\documentclass[12pt]{article}
\usepackage{amsmath,amssymb}
\usepackage{graphicx}
\newcommand{\eqdef}{\stackrel{\text{def}}{=}}
\newcommand{\n}{\nonumber\\}
\newcommand{\bm}{\boldsymbol}
\newcommand{\ignore}[1]{}
\numberwithin{equation}{section}
\newcommand{\Romannumeral}[1]{\uppercase\expandafter{\romannumeral#1}}

\newcommand{\II}{\text{\Romannumeral{2}}}
\newcommand{\U}{\text{\bf U}}

\allowdisplaybreaks[4]

\begin{document}

\baselineskip=20pt

\newcommand{\preprint}{
    \begin{flushright}\normalsize \sf
     DPSU-14-1\\
   \end{flushright}}
\newcommand{\Title}[1]{{\baselineskip=26pt
   \begin{center} \Large \bf #1 \\ \ \\ \end{center}}}
\newcommand{\Author}{\begin{center}
   \large \bf Satoru Odake and Ryu Sasaki \end{center}}
\newcommand{\Address}{\begin{center}
     Department of Physics, Shinshu University,\\
     Matsumoto 390-8621, Japan
    \end{center}}
\newcommand{\Accepted}[1]{\begin{center}
   {\large \sf #1}\\ \vspace{1mm}{\small \sf Accepted for Publication}
   \end{center}}

\preprint
\thispagestyle{empty}

\Title{Solvable Discrete Quantum Mechanics: $q$-Orthogonal Polynomials
with $|q|=1$ and Quantum Dilogarithm}

\Author

\Address
\vspace{1cm}

\begin{abstract}
Several kinds of $q$-orthogonal polynomials with $|q|=1$ are constructed
as the main parts of the eigenfunctions of new solvable discrete quantum
mechanical systems.
Their orthogonality weight functions consist of quantum dilogarithm functions,
which are a natural extension of the Euler gamma functions and the $q$-gamma
functions ($q$-shifted factorials). The dimensions of the orthogonal spaces
are finite. These $q$-orthogonal polynomials are expressed in terms of the
Askey-Wilson polynomials and their certain limit forms.
\end{abstract}

\section{Introduction}
\label{sec:intro}

Several years ago the present authors published ``Unified theory of exactly and
quasiexactly solvable `discrete' quantum mechanics.\,I.\,Formalism,''\cite{os14}
(to be referred to as \U, and its equations are cited as (\U2.3) etc).
It provided a unified framework for the known solvable discrete quantum
mechanical systems \cite{os24} with pure imaginary shifts (idQM) and with
real shifts (rdQM). The Wilson and the Askey-Wilson polynomials
\cite{askey,ismail,koeswart,os13} are the main parts of the eigenpolynomials
of solvable idQM, whereas the (dual) ($q$-)Hahn and the ($q$-)Racah
polynomials \cite{askey,ismail,koeswart,os12} belong to rdQM.
The unified theory \U\ provides not only a synthesised explanation
of the two sufficient conditions of exact solvability, {\em i.e.\/} shape
invariance \cite{genden} and the closure relations \cite{os7}, but also
a systematic method to introduce new solvable and quasi-exactly solvable
systems based on sinusoidal coordinates \cite{os7}.

In this paper we present several new solvable systems in idQM and
$q$-orthogonal polynomials as the main parts of their eigenfunctions.
They correspond to the choices of the sinusoidal coordinates
$\eta(x)=\cosh x,e^{\pm x},\sinh x$, whereas that of the Askey-Wilson system
is $\cos x$. The new system  with $\eta(x)=\cosh x$ is obtained from the
Askey-Wilson system by the replacement $x^{\text{AW}}\to-ix$
\eqref{okikae(vii)}, which induces the change of the $q$-parameter 
$0<q^{\text{AW}}<1 \to q=e^{-i\gamma}$, $0<\gamma<\pi$, $|q|=1$.
This gives rise to $q$-orthogonal polynomials with $|q|=1$ as the main
parts of the eigenfunctions. The orthogonality weight functions consist of
quantum dilogarithm functions \cite{qdilog}. The weight functions for the
orthogonal polynomials of the known solvable idQM are the Euler gamma
functions $\Gamma(z)$ for $\eta(x)=x,x^2$ (the continuous Hahn and the
Wilson polynomials etc) and the $q$-gamma functions
$\Gamma_q(z)=(1-q)^{1-z}(q;q)_{\infty}(q^z;q)_{\infty}^{-1}$ or the
$q$-shifted factorials $(a;q)_{\infty}$ for $\eta(x)=\cos x, \sin x$
(the Askey-Wilson polynomials etc).
The double gamma functions or the quantum dilogarithms are naturally expected
for $\eta(x)=\cosh x,e^{\pm x},\sinh x$, having $|q|=1$.
For example, in the XXZ spin chain with spin one-half which has quantum
algebra symmetry, various quantities in the antiferromagnetic regime
$\Delta<-1$ ($-1<q<0$) are described by using the $q$-shifted factorials
$(a;q)_{\infty}$, and those in the gapless regime $|\Delta|\leq 1$ ($|q|=1$)
are described by using the double sine function \cite{qdilog2}.
The double gamma function and the double sine function are known for more
than a century \cite{barnes,shintani,kurokawa} and they are closely related
to quantum dilogarithm functions \cite{qdilog,qdilog2}.

Another way to understand the appearance of quantum dilogarithm in comparison
with the ($q$-)gamma functions is their {\em functional equations\/}.
Within the framework of idQM, the orthogonality weight function is obtained
as the square of the groundstate wavefunction $\phi_0(x)$, which is the
zero mode of the operator $\mathcal{A}$, \eqref{Aphi0=0} of the factorised
Hamiltonian \eqref{H}. As a building block of the groundstate wavefunction,
we need a solution $F(x)$ of the following functional equation
\begin{equation*}
  \frac{F(x-i\frac{\gamma}{2})}{F(x+i\frac{\gamma}{2})}
  =\frac{v(x+i\frac{\gamma}{2})}{v^*(x-i\frac{\gamma}{2})},
\end{equation*}
where $v(x)$ is a main factor of the potential function $V(x)$.
For the eight types of the sinusoidal coordinates introduced in
(\U A.1)--(\U A.8) and listed in \eqref{etalist_os14}, the main factors
of the potentials are
\begin{equation*}
  \text{(\romannumeral1)--(\romannumeral2)}\ \ v(x)=a+ix,\quad
  \text{(\romannumeral3)--(\romannumeral4)}\ \ v(x)=1-ae^{ix},\quad
  \text{(\romannumeral5)--(\romannumeral8)}\ \ v(x)=1+ae^x.
\end{equation*}
If we find an $F_1(x)$ which solves
$\frac{F_1(x-i\frac{\gamma}{2})}{F_1(x+i\frac{\gamma}{2})}
=v(x+i\frac{\gamma}{2})$,
then $F(x)=F_1(x)F_1^*(x)\bigl(=F^*(x)\bigr)$ solves the above equation.
For the above $v(x)$, we have
\begin{align*}
  &\text{(\romannumeral1)--(\romannumeral2)}\ \ F_1(x)=\Gamma(a+ix),
  \ \gamma=1,\quad
  \text{(\romannumeral3)--(\romannumeral4)}
  \ \ F_1(x)=(ae^{ix};e^{\gamma})_{\infty}^{-1},\\
  &\text{(\romannumeral5)--(\romannumeral8)}
  \ \ F_1(x)=\Phi_{\frac{\gamma}{2}}(x+i\tfrac{\gamma}{2}+\log a),
\end{align*}
where $\Phi_{\gamma}(z)$ is the quantum dilogarithm introduced in Appendix
\ref{app:qdilog} and satisfies the functional equation \eqref{funcrel}.

The present paper is organised as follows.
The key components of the foundation paper \U\ are recapitulated in section two.
Starting from the essence of the discrete quantum mechanics with pure
imaginary shifts (idQM) in \S\,\ref{sec:idQM}, the construction method of
the solvable potentials based on sinusoidal coordinates is summarised in
\S\,\ref{sec:os14}. A brief note on the hermiticity of the resulting
Hamiltonians is given in \S\,\ref{sec:hermite}.
Section three is the main part of the paper, providing the details of
the four new idQM systems; the factorised potential functions, the
eigenpolynomials, the ground state eigenfunctions or the orthogonality
weight functions and verifications of the hermiticity.
The representative system having the sinusoidal coordinate $\eta(x)=\cosh x$
is discussed in detail in \S\,\ref{sec:coshx}. Comments on the `roots of unity
$q$' cases are given in \S\,\ref{sec:coshx,comment}. 
The cases of $\eta(x)=e^{\pm x}$ are discussed briefly  in \S\,\ref{sec:e^x}.
The system with the sinusoidal coordinate $\eta(x)=\sinh x$ is discussed in
\S\,\ref{sec:sinhx}. Comments on the closure relation and shape invariance of
the new solvable systems are given in \S\,\ref{sec:cr} and \S\,\ref{sec:si},
respectively.
The normalisation constants are discussed in \S\,\,\ref{sec:hn}.
By construction the discrete quantum mechanics reduces to the ordinary
quantum mechanics in certain limits. In \S\,\ref{sec:limit} the new solvable
idQM systems derived in \S\,\ref{sec:coshx}--\S\,\ref{sec:sinhx} are shown to
reduce to the known solvable systems with the Morse, hyperbolic
Darboux-P\"{o}schl-Teller and hyperbolic symmetric top $\II$ potentials
\cite{infhul,susyqm,os28,os29}, at the levels of the Hamiltonian, the
ground state wave function and the eigenpolynomials. Other limiting
properties are discussed in \S\,\ref{sec:limit2}.
The final section is for a summary and comments.
In Appendix A, various data of the solvable idQM systems corresponding to
the eight choices of the sinusoidal coordinates \eqref{etalist_os14} are
described.
In Appendix B, a brief summary of quantum dilogarithm functions is presented.

\section{Solvable Discrete Quantum Mechanics}
\label{sec:formalism}

In order to set the stage and to introduce proper notation we recapitulate
the essence of the discrete quantum mechanics with pure imaginary shifts
(idQM) \cite{os24,os13} and the method to construct exactly solvable systems
presented in \U.

\subsection{Discrete quantum mechanics with pure imaginary shifts}
\label{sec:idQM}

In idQM \cite{os24,os13} the dynamical variables are the real coordinate
$x$ ($x_1<x<x_2$) and the conjugate momentum $p=-i\partial_x$.
The Hamiltonian is positive semi-definite due to the factorised form:
\begin{align}
  &\mathcal{H}\eqdef\sqrt{V(x)}\,e^{\gamma p}\sqrt{V^*(x)}
  +\!\sqrt{V^*(x)}\,e^{-\gamma p}\sqrt{V(x)}-V(x)-V^*(x)
  =\mathcal{A}^{\dagger}\mathcal{A},
  \label{H}\\
  &\mathcal{A}\eqdef i\bigl(e^{\frac{\gamma}{2}p}\sqrt{V^*(x)}
  -e^{-\frac{\gamma}{2}p}\sqrt{V(x)}\,\bigr),\quad
  \mathcal{A}^{\dagger}\eqdef-i\bigl(\sqrt{V(x)}\,e^{\frac{\gamma}{2}p}
  -\sqrt{V^*(x)}\,e^{-\frac{\gamma}{2}p}\bigr).
  \label{A}
\end{align}
Here the potential function $V(x)$ is an analytic function of $x$ and
$\gamma$ is a real constant.
The $*$-operation on an analytic function $f(x)=\sum_na_nx^n$
($a_n\in\mathbb{C}$) is defined by $f^*(x)\eqdef f(x^*)^*=\sum_na_n^*x^n$,
in which $a_n^*$ is the complex conjugation of $a_n$.
Obviously $f^{**}(x)=f(x)$ and $f(x)^*=f^*(x^*)$.
The inner product of two functions $f$ and $g$ is the integral of the
product of two analytic functions $(f,g)\eqdef\int_{x_1}^{x_2}dxf^*(x)g(x)$.
The Schr\"{o}dinger equation
\begin{align}
  &\mathcal{H}\phi_n(x)=\mathcal{E}_n\phi_n(x)\quad(n=0,1,\ldots),\quad
  0=\mathcal{E}_0<\mathcal{E}_1<\cdots,
  \label{Scheq}
\end{align}
is an analytic difference equation with  pure imaginary shifts
($e^{\pm\gamma p}f(x)=f(x\mp i\gamma)$).
The orthogonality relation reads
\begin{equation}
  (\phi_n,\phi_m)=h_n\delta_{nm}\quad(0<h_n<\infty),
  \label{ortho}
\end{equation}
and the number of the eigenstates ($h_n<\infty$) is finite or infinite.
The eigenfunction $\phi_n(x)$ can be chosen `real', $\phi_n^*(x)=\phi_n(x)$,
and we follow this convention.
The groundstate wavefunction $\phi_0(x)$ is the zero mode of the operator
$\mathcal{A}$, $\mathcal{A}\phi_0(x)=0$, namely
\begin{equation}
  \sqrt{V^*(x-i\tfrac{\gamma}{2})}\,\phi_0(x-i\tfrac{\gamma}{2})
  =\sqrt{V(x+i\tfrac{\gamma}{2})}\,\phi_0(x+i\tfrac{\gamma}{2}).
  \label{Aphi0=0}
\end{equation}

The higher eigenfunctions have factorised forms,
\begin{equation}
  \phi_n(x)=\phi_0(x)P_n\bigl(\eta(x)\bigr),
  \label{phin_form}
\end{equation}
in which $P_n(\eta)$ is a polynomial of degree $n$ in the sinusoidal coordinate
$\eta=\eta(x)$.
This form implies that $P_n(\eta)$'s ($n=0,1,\ldots$) are orthogonal
polynomials (in the ordinary sense) with the weight function $\phi_0(x)^2$.
The Schr\"{o}dinger equation \eqref{Scheq} becomes
\begin{equation}
  \widetilde{\mathcal{H}}P_n\bigl(\eta(x)\bigr)
  =\mathcal{E}_nP_n\bigl(\eta(x)\bigr),
  \label{HtPn}
\end{equation}
where the similarity transformed Hamiltonian $\widetilde{\mathcal{H}}$ is
square root free
\begin{equation}
  \widetilde{\mathcal{H}}\eqdef\phi_0(x)^{-1}\circ\mathcal{H}\circ\phi_0(x)
  =V(x)(e^{\gamma p}-1)+V^*(x)(e^{-\gamma p}-1).
  \label{Ht}
\end{equation}
In this formalism, the orthogonal polynomials $\{P_n(\eta)\}$ and their weight
function $\phi_0(x)^2$ are determined by the given form of the potential
function $V(x)$.
With abuse of language, we call \eqref{HtPn} a polynomial equation, which is
in fact a  difference equation for the polynomial $P_n(\eta(x))$.

\subsection{Construction of solvable potentials}
\label{sec:os14}

In \U\ a systematic method to obtain exactly and quasi-exactly solvable systems
was presented for discrete quantum mechanics.
Here we summarise it for idQM, focusing on the exactly solvable case.

As seen above, the Schr\"{o}dinger equation \eqref{Scheq} becomes the
polynomial equation \eqref{HtPn} by the similarity transformation with
respect to the groundstate wavefunction $\phi_0(x)$.
The method presented in \U\ is reversing this order:
\begin{enumerate}
\setlength{\parskip}{0cm} 
\setlength{\itemsep}{0cm} 
\item[1)]
  Construct exactly solvable polynomial equation,
  $\widetilde{\mathcal{H}}P_n\bigl(\eta(x)\bigr)
  =\mathcal{E}_nP_n\bigl(\eta(x)\bigr)$,
  based on the sinusoidal coordinate $\eta(x)$.
\item[2)]
  Find the groundstate wavefunction $\phi_0(x)$, which satisfies
  \eqref{Aphi0=0} and ensures the hermiticity of the Hamiltonian.
\item[3)]
  By the similarity transformation
  $\mathcal{H}=\phi_0(x)\circ\widetilde{\mathcal{H}}\circ\phi_0(x)^{-1}$,
  exactly solvable quantum system is obtained, 
  $\mathcal{H}\phi_n(x)=\mathcal{E}_n\phi_n(x)$,
  $\phi_n(x)=\phi_0(x)P_n\bigl(\eta(x)\bigr)$. 
\end{enumerate}

\subsubsection{sinusoidal coordinate}
\label{sec:eta}

The sinusoidal coordinate $\eta(x)$ for idQM is defined as a `real'
analytic ($\eta^*(x)=\eta(x)$) function of $x$ satisfying the following
(\U 2.18), (\U 2.20):
\begin{equation}
  \left\{
  \begin{array}{ll}
  0)&
  \text{$\eta(x)$ and $x$ are one-to-one for $x_1<x<x_2$}\\[2pt]
  1)&
  \eta(x-i\gamma)+\eta(x+i\gamma)=(2+r_1^{(1)})\eta(x)+r_{-1}^{(2)}\\[4pt]
  2)&
  \bigl(\eta(x-i\gamma)-\eta(0)\bigr)\bigl(\eta(x+i\gamma)-\eta(0)\bigr)
  =\bigl(\eta(x)-\eta(-i\gamma)\bigr)\bigl(\eta(x)-\eta(i\gamma)\bigr)
  \end{array}\right..
  \label{eta_os14}
\end{equation}
Here $r_1^{(1)}$ and $r_{-1}^{(2)}$ are real parameters and we assume
$r_1^{(1)}>-4$ and $\eta(x)\neq\eta(x-i\gamma)\neq\eta(x+i\gamma)\neq\eta(x)$.
The two conditions 1) \& 2) imply that any symmetric polynomial in
$\eta(x-i\gamma)$ and $\eta(x+i\gamma)$ is expressed as a polynomial in
$\eta(x)$.
In this paper we do not require the condition $\eta(0)=0$, which was imposed
in \U\ due to a unified presentation for idQM and rdQM.
An affine transformed one $\eta^{\text{new}}(x)=a\eta(x)+b$ ($a,b$:
constants, $a\neq 0$) works as a sinusoidal coordinate, and the argument
of exact solvability does not change.

\subsubsection{potential function}
\label{sec:V}

We assume that the potential function $V(x)$ has the following form
(\U 2.30)--(\U 2.31):
\begin{align}
  V(x)&=\frac{\widetilde{V}(x)}
  {\bigl(\eta(x-i\gamma)-\eta(x)\bigr)
  \bigl(\eta(x-i\gamma)-\eta(x+i\gamma)\bigr)}\,,
  \label{V=Vt/etaeta}\\
  \widetilde{V}(x)&=\sum_{\genfrac{}{}{0pt}{}{k,l\geq 0}{k+l\leq 2}}
  v_{k,l}\,\eta(x)^k\eta(x-i\gamma)^l,
  \label{Vt}
\end{align}
where $v_{k,l}$ are real constants ($\sum_{k+l=2}v_{k,l}^2\neq 0$).
Due to the properties \eqref{eta_os14}, the number of independent real
parameters among $\{v_{k,l}\,(k+l\leq 2)\}$ (concerning the functional form
of $\widetilde{V}(x)$) is five and one of them corresponds to an overall
normalisation.
For example, we can take $v_{k,l}$ $(k+l\leq 2,\ l=0,1$) as independent
parameters, since $v_{0,2}$ is redundant as
\begin{align}
  \eta(x-i\gamma)^2
  &=(2+r_1^{(1)})\eta(x)\eta(x-i\gamma)-\eta(x)^2
  +r_{-1}^{(2)}\bigl(\eta(x)+\eta(x-i\gamma)\bigr)\n
  &\quad-\eta(-i\gamma)\eta(i\gamma)+\eta(0)\bigl(\eta(0)-r_{-1}^{(2)}\bigr).
  \label{eta(x-ig)^2}
\end{align}

\subsubsection{polynomial equation}
\label{sec:polyeq}

Our starting point is the similarity transformed Hamiltonian
$\widetilde{\mathcal{H}}$ :
\begin{equation*}
  \widetilde{\mathcal{H}}=V(x)(e^{\gamma p}-1)+V^*(x)(e^{-\gamma p}-1).
\end{equation*}
This acts on $\eta(x)^n$ ($n=0,1,\ldots$) as (\U 2.36)
\begin{equation}
  \widetilde{\mathcal{H}}\eta(x)^n
  =\sum_{m=0}^n\eta(x)^m\widetilde{\mathcal{H}}_{m,n}^{\eta},
  \label{Htact}
\end{equation}
where the explicit forms of the real matrix elements
$\widetilde{\mathcal{H}}^{\eta}_{m,n}$ are given in Appendix \ref{app:Htmn}.
The polynomial space
$\mathcal{V}_n\eqdef\text{Span}\bigl[1,\eta(x),\ldots,\eta(x)^n\bigr]$ is
invariant under the action of $\widetilde{\mathcal{H}}$ :
\begin{equation}
  \widetilde{\mathcal{H}}\mathcal{V}_n\subseteq\mathcal{V}_n.
\end{equation}
The matrix $(\widetilde{\mathcal{H}}^{\eta}_{m,n})$ is an upper triangular
matrix. The solution of \eqref{HtPn} is given explicitly in a determinant
form (\U 3.2),
\begin{equation}
  \mathcal{E}_n=\widetilde{H}^{\eta}_{n,n},\quad
  P_n\bigl(\eta(x)\bigr)\propto
  \begin{vmatrix}
    1&\eta(x)&\eta(x)^2&\cdots&\eta(x)^n\\[2pt]
    \,\mathcal{E}_0-\mathcal{E}_n&\widetilde{H}^{\eta}_{0,1}&
    \widetilde{H}^{\eta}_{0,2}&\cdots&\widetilde{H}^{\eta}_{0,n}\\[2pt]
    &\mathcal{E}_1-\mathcal{E}_n&\widetilde{H}^{\eta}_{1,2}&
    \cdots&\widetilde{H}^{\eta}_{1,n}\\[2pt]
    &&\ddots&\ddots&\vdots\\[2pt]
    \text{\LARGE $0$}&&&\mathcal{E}_{n-1}-\mathcal{E}_n&
    \widetilde{H}^{\eta}_{n-1,n}\,
  \end{vmatrix}.
  \label{Pndet}
\end{equation}
Note that the determinant in \eqref{Pndet} is `real' ($f(x)=|\cdots|$,
$f^*(x)=f(x)$) since each component of the matrix is `real'.

\subsubsection{exactly solvable Hamiltonian}
\label{sec:}

The groundstate wavefunction $\phi_0(x)$ is required to obey \eqref{Aphi0=0}
and it should be square integrable.
Contrary to the ordinary quantum mechanics in which the groundstate
wavefunction satisfies a first order differential equation, the solution of
the difference equation \eqref{Aphi0=0} is not unique.
However, the groundstate wavefunction is uniquely determined by the hermiticity
requirement of the Hamiltonian up to an overall normalisation, as will be
shown shortly in the subsequent subsection \S\,\ref{sec:hermite}.

Once the desired groundstate wavefunction $\phi_0(x)$ is obtained,
the inverse similarity transformation of $\widetilde{\mathcal{H}}$
with respect to $\phi_0(x)$ gives an exactly solvable idQM,
\begin{equation*}
  \mathcal{H}=\phi_0(x)\circ\widetilde{\mathcal{H}}\circ\phi_0(x)^{-1},\quad
  \mathcal{H}\phi_n(x)=\mathcal{E}_n\phi_n(x),\quad
  \phi_n(x)=\phi_0(x)P_n\bigl(\eta(x)\bigr).
\end{equation*}
The existence of the groundstate wavefunction $\phi_0(x)$ and the hermiticity
of the Hamiltonian $\mathcal{H}$ restrict the range of parameters
$\{v_{k,l}\}$. Depending on the choice of the sinusoidal coordinate
$\eta(x)$ \eqref{etalist_os14}, the number of square-integrable eigenstates
$\phi_n(x)$ is finite or infinite.

As explained in Introduction, it is very important to {\em factorise the
potential function\/} $V(x)$ for solving the zero mode equation
\eqref{Aphi0=0} for $\phi_0(x)$.
Then \eqref{Aphi0=0} can be solved for each factor of $V(x)$ and the
multiplication of each factor solution would give $\phi_0(x)$.
This type of solution method for $\phi_0(x)$ is characteristic to difference
equations and it is very different from the corresponding situation for
the differential equations.
The factorisation naturally introduces {\em new parametrisation of the
potential function\/} $V(x)$, which provides inherited parametrisation
of the weight function $\phi_0(x)^2$ and the polynomials $\{P_n(\eta)\}$.

The present method also covers the known solvable potentials in idQM;
for example, those corresponding to the Wilson and the Askey-Wilson
polynomials. In these cases, the natural parameters are not the original
$\{v_{k,l}\}$ \eqref{Vt} in \S\,\ref{sec:V} but those obtained by factorisation.
As for the new examples of solvable idQM in \S\,\ref{sec:example} the
situation is the same.
The starting point will be the factorised potential functions \eqref{V(vii)},
\eqref{V(v)}, \eqref{V(vi)} and \eqref{V(viii)}.
The relation among the original parameters $\{v_{k,l}\}$ \eqref{Vt} and
the factorisation parameters are given in \eqref{vjk-aj}.

\subsection{Hermiticity}
\label{sec:hermite}

The hermiticity of the Hamiltonian ($\mathcal{H}=p^2+U(z)$) in ordinary
quantum mechanics (oQM) is simple but that of idQM is involved
\cite{os14,os13,os27}.
Here we review the hermiticity of the Hamiltonian \eqref{H} with the
eigenfunctions \eqref{phin_form}.
It should be stressed that the groundstate wavefunction $\phi_0(x)$ contains
square roots but the weight function $\phi_0(x)^2$ is square root free.

Let us consider the functions of the form
$f(x)=\phi_0(x)\mathcal{P}\bigl(\eta(x)\bigr)$, where $\mathcal{P}(\eta)$
is a polynomial  in $\eta$ and $\mathcal{P}^*(\eta)=\mathcal{P}(\eta)$.
For two functions $f_1(x)=\phi_0(x)\mathcal{P}_1\bigl(\eta(x)\bigr)$ and
$f_2(x)=\phi_0(x)\mathcal{P}_2\bigl(\eta(x)\bigr)$
($\text{deg}\,\mathcal{P}_1=n_1$, $\text{deg}\,\mathcal{P}_2=n_2$),
the hermiticity $(f_1,\mathcal{H}f_2)=(\mathcal{H}f_1,f_2)$ is realised
if the following quantity vanishes \cite{os27}:
\begin{align}
  &\quad\int_{-\frac{\gamma}{2}}^{\frac{\gamma}{2}} 
  dy\bigl(G(x_2+iy)-G^*(x_2-iy)\bigr)
  -\int_{-\frac{\gamma}{2}}^{\frac{\gamma}{2}} 
  dy\bigl(G(x_1+iy)-G^*(x_1-iy)\bigr)\n
  &=2\pi\frac{\gamma}{|\gamma|}\!\sum_{x_0:\text{pole in $D_{\gamma}$}}
  \!\!\!\text{Res}_{x=x_0}\bigl(G(x)-G^*(x)\bigr),
  \label{H=Hd}
\end{align}
where $G(x)$ and $D_{\gamma}$ are
\begin{align}
  &G(x)=V(x+i\tfrac{\gamma}{2})\phi_0(x+i\tfrac{\gamma}{2})^2
  \,\mathcal{P}_1\bigl(\eta(x+i\tfrac{\gamma}{2})\bigr)
  \mathcal{P}_2\bigl(\eta(x-i\tfrac{\gamma}{2})\bigr),\n
  &\ \ \Bigl(\Rightarrow
  G^*(x)=V(x+i\tfrac{\gamma}{2})\phi_0(x+i\tfrac{\gamma}{2})^2
  \,\mathcal{P}_1\bigl(\eta(x-i\tfrac{\gamma}{2})\bigr)
  \mathcal{P}_2\bigl(\eta(x+i\tfrac{\gamma}{2})\bigr)\Bigr),
  \label{Gdef}\\
  &D_{\gamma}\eqdef\bigl\{x\in\mathbb{C}\bigm|x_1\leq\text{Re}\,x\leq x_2,
  |\text{Im}\,x|\leq\tfrac12|\gamma|\bigr\}.
  \label{Dgamma}
\end{align}
In the next section we will verify  the condition \eqref{H=Hd} for concrete
examples.
The degree of polynomial $\mathcal{P}(\eta)$ may have upper limit due to
square integrability.

\section{New Examples of Exactly Solvable idQM}
\label{sec:example}

The following eight sinusoidal coordinates are presented in Appendix A of
\U\ (in which affine transformed $\eta(x)$ are used for
(\romannumeral3) and (\romannumeral5)--(\romannumeral7)) :
\begin{alignat}{3}
  \text{(\romannumeral1)}:&\quad&\eta(x)&=x,
  &\quad -\infty<\,&x<\infty,\n[-3pt]
  \text{(\romannumeral2)}:&\quad&\eta(x)&=x^2,
  &\quad 0<\,&x<\infty,\n[-3pt]
  \text{(\romannumeral3)}:&\quad&\eta(x)&=\cos x,
  &\quad 0<\,&x<\pi,\n[-3pt]
  \text{(\romannumeral4)}:&\quad&\eta(x)&=\sin x,
  &\quad -\tfrac{\pi}{2}<\,&x<\tfrac{\pi}{2},\n[-3pt]
  \text{(\romannumeral5)}:&\quad&\eta(x)&=e^{-x},
  &\quad -\infty<\,&x<\infty,
  \label{etalist_os14}\\[-3pt]
  \text{(\romannumeral6)}:&\quad&\eta(x)&=e^x,
  &\quad -\infty<\,&x<\infty,\n[-3pt]
  \text{(\romannumeral7)}:&\quad&\eta(x)&=\cosh x,
  &\quad 0<\,&x<\infty,\n[-3pt]
  \text{(\romannumeral8)}:&\quad&\eta(x)&=\sinh x,
  &\quad -\infty<\,&x<\infty.\nonumber
\end{alignat}
We take $\gamma$ as
\begin{equation}
  \text{(\romannumeral1)--(\romannumeral2)}: \gamma=1,\quad
  \text{(\romannumeral3)--(\romannumeral4)}: \gamma<0,\quad
  \text{(\romannumeral5)--(\romannumeral8)}: 0<\gamma<\pi.
  \label{gamma_range}
\end{equation}
These eight sinusoidal coordinates satisfy \eqref{eta_os14} and
\eqref{etahalf} with
\begin{alignat}{2} 
  \text{(\romannumeral1)}&: r^{(1)}_1=0,\ r^{(2)}_{-1}=0, &\quad 
  \text{(\romannumeral2)}&: r^{(1)}_1=0,\ r^{(2)}_{-1}=-2, \n
  \text{(\romannumeral3)--(\romannumeral4)}&: 
  r^{(1)}_1=(e^{\frac{\gamma}{2}}-e^{-\frac{\gamma}{2}})^2,
  \ r^{(2)}_{-1}=0, &\quad 
 \text{(\romannumeral5)--(\romannumeral8)}&: 
  r^{(1)}_1=(e^{i\frac{\gamma}{2}}-e^{-i\frac{\gamma}{2}})^2,\ r^{(2)}_{-1}=0. 
  \label{r11}
\end{alignat}

The cases (\romannumeral1)--(\romannumeral3) give the {\em well-known quantum
systems with infinitely many discrete eigenstates\/}, whose eigenfunctions
are described by (\romannumeral1) continuous Hahn, (\romannumeral2) Wilson
and (\romannumeral3) Askey-Wilson polynomials.
In the case (\romannumeral3) the parameter $q=e^{\gamma}$ of the Askey-Wilson
polynomials satisfies $0<q<1$.
The case (\romannumeral4) is nothing but the case (\romannumeral3) with
the shift of the coordinate,
$x^{\text{(\romannumeral4)}}=x^{\text{(\romannumeral3)}}-\frac{\pi}{2}$.

The {\em new examples of solvable idQM\/} are obtained by the sinusoidal
coordinates (\romannumeral5)--(\romannumeral8). We will explain
$\eta(x)=\cosh x$ (\romannumeral7) case in detail and others briefly.
These systems have {\em finitely many discrete eigenstates\/}.
The explicit forms of the eigenpolynomials \eqref{Pndet} are listed in
\eqref{exPndet}. These eigenpolynomials are given by the Askey-Wilson
polynomial with the parameter $q=e^{-i\gamma}$, $|q|=1$ and its limiting forms.
Their groundstate wavefunctions are described by the quantum dilogarithm.
In other words, the present research presents four kinds of (finitely many)
$q$-orthogonal polynomials with $|q|=1$ and having quantum dilogarithm
functions as their weight functions.
 
For these eight sinusoidal coordinates, the potential function $V(x)$
\eqref{V=Vt/etaeta}--\eqref{Vt} in the original parameters $\{v_{k,l}\}$
are listed in \eqref{exV} and the {\em factorised forms\/} are listed in
\eqref{Vtfactor}.

Let us denote the factorisation parameters as
$\bm{\lambda}=(\lambda_1,\lambda_2,\ldots)$.
We write the parameter dependence explicitly, as $\mathcal{H}(\bm{\lambda})$,
$\mathcal{A}(\bm{\lambda})$, $\mathcal{E}_n(\bm{\lambda})$,
$\phi_n(x;\bm{\lambda})$, etc.

\subsection{$\bm{\eta(x)=\cosh x}$}
\label{sec:coshx}

Here we consider the case (\romannumeral7) $\eta(x)=\cosh x$, $0<x<\infty$,
$0<\gamma<\pi$, in detail.

\subsubsection{factorised potential function}
\label{sec:coshx,V}

For this choice of the sinusoidal coordinate, the potential function $V(x)$
in the original parameter $\{v_{k,l}\}$ \eqref{V=Vt/etaeta}--\eqref{Vt} is
listed in \eqref{exV}. The factorised form is shown in \eqref{Vtfactor}.
By choosing the overall scaling factor
$A=\frac{\sin\frac{\gamma}{2}\sin\gamma}{|a_1a_2|}$,
we adopt the factorised potential:
\begin{equation}
  V(x;\bm{\lambda})
  =e^{i\pi}e^{-i\frac{\gamma}{2}}\frac{a_1^*a_2^*}{|a_1a_2|}
  \frac{\prod_{j=1}^2(1+a_je^x)(1+a_j^{*\,-1}e^x)}
  {(e^{2x}-1)(e^{-i\gamma}e^{2x}-1)},
  \label{V(vii)}
\end{equation}
where $e^{i\pi}$ indicates $\sqrt{e^{i\pi}}=e^{\frac{i}{2}\pi}$ in
$\sqrt{V(x;\bm{\lambda})}$.
Here the relation between $\bm{\lambda}=(\lambda_1,\lambda_2)$ and
$a_j\in\mathbb{C}_{\neq 0}$ ($j=1,2$) is
\begin{equation}
  a_j\eqdef e^{-i\gamma\lambda_j},\quad\lambda_j\eqdef\alpha_j+i\beta_j,\quad
  -\pi<\gamma\alpha_j\leq\pi,\quad\beta_j\in\mathbb{R},\quad
  \alpha\eqdef\alpha_1+\alpha_2.
  \label{a_j}
\end{equation}
Note that $\frac{a_1^*a_2^*}{|a_1a_2|}=e^{i\gamma\alpha}$ and
\begin{equation}
  V^*(x;\bm{\lambda})=V(-x;\bm{\lambda}).
  \label{V*(-x)(vii)}
\end{equation}
This potential function has four real parameters $\alpha_1$, $\alpha_2$,
$\beta_1$ and $\beta_2$ in addition to $\gamma$.

\subsubsection{eigenpolynomials}
\label{sec:coshx,Pn}

The determinant formula for the eigenpolynomial \eqref{Pndet} gives
\eqref{exPndet} in the new parameters. So we take $P_n(\eta;\bm{\lambda})$
and $\mathcal{E}_n(\bm{\lambda})$ as
\begin{align}
  P_n(\eta;\bm{\lambda})&
  =e^{-i\frac{\pi}{2}n}e^{i\gamma\frac34n(n-1)}e^{i\gamma\alpha n}\,
  p_n(\eta;-a_1,-a_2,-a_1^{*\,-1},-a_2^{*\,-1}|e^{-i\gamma}),
  \label{Pn(vii)}\\
  \mathcal{E}_n(\bm{\lambda})&
  =4\sin\tfrac{\gamma}{2}n\,\sin\tfrac{\gamma}{2}(n-1+2\alpha),
  \label{En(vii)}
\end{align}
where $p_n$ is the Askey-Wilson polynomial \eqref{AWpoly}.
By using \eqref{AWpoly} and \eqref{pnprop1}--\eqref{pnprop2}, we can show
the reality of the polynomial,
\begin{equation}
  P_n^*(\eta;\bm{\lambda})=P_n(\eta;\bm{\lambda}). 
  \label{Pn*}
\end{equation}

Let us prove that $P_n(\eta;\bm{\lambda})$ \eqref{Pn(vii)} and
$\mathcal{E}_n(\bm{\lambda})$ \eqref{En(vii)} satisfy \eqref{HtPn}.
We know that the Askey-Wilson polynomial satisfies \eqref{HtPn},
\begin{align}
  &\quad\frac{\prod_{j=1}^4(1-a_j^{\text{AW}}e^{ix^{\text{AW}}})}
  {(1-e^{2ix^{\text{AW}}})(1-e^{\gamma^{\text{AW}}}e^{2ix^{\text{AW}}})}
  \Bigl(p_n\bigl(\cos(x^{\text{AW}}-i\gamma^{\text{AW}});
  a_1^{\text{AW}},a_2^{\text{AW}},a_3^{\text{AW}},a_4^{\text{AW}}
  |e^{\gamma^{\text{AW}}}\bigr)\n
  &\qquad\qquad\qquad\qquad\qquad\qquad\quad
  -p_n(\cos x^{\text{AW}};a_1^{\text{AW}},a_2^{\text{AW}},
  a_3^{\text{AW}},a_4^{\text{AW}}|e^{\gamma^{\text{AW}}})\Bigr)\n
  &+\frac{\prod_{j=1}^4(1-a_j^{\text{AW}}e^{-ix^{\text{AW}}})}
  {(1-e^{-2ix^{\text{AW}}})(1-e^{\gamma^{\text{AW}}}e^{-2ix^{\text{AW}}})}
  \Bigl(p_n\bigl(\cos(x^{\text{AW}}+i\gamma^{\text{AW}});
  a_1^{\text{AW}},a_2^{\text{AW}},a_3^{\text{AW}},a_4^{\text{AW}}
  |e^{\gamma^{\text{AW}}}\bigr)
  \label{HtPnAW}\\
  &\qquad\qquad\qquad\qquad\qquad\qquad\qquad
  -p_n(\cos x^{\text{AW}};a_1^{\text{AW}},a_2^{\text{AW}},
  a_3^{\text{AW}},a_4^{\text{AW}}|e^{\gamma^{\text{AW}}})\Bigr)\n
  &=(e^{-\gamma^{\text{AW}}n}-1)(1-a_1^{\text{AW}}a_2^{\text{AW}}
  a_3^{\text{AW}}a_4^{\text{AW}}e^{\gamma^{\text{AW}}(n-1)})
  p_n(\cos x^{\text{AW}};a_1^{\text{AW}},a_2^{\text{AW}},
  a_3^{\text{AW}},a_4^{\text{AW}}|e^{\gamma^{\text{AW}}}).
  \nonumber
\end{align}
We stress that this equation holds for any complex values of the parameters
$a_j^{\text{AW}}$ and $\gamma^{\text{AW}}$ and the coordinate $x^{\text{AW}}$
(except for the zeros of denominators).
By substituting
\begin{equation}
  x^{\text{AW}}\to-ix,\quad\gamma^{\text{AW}}\to-i\gamma,\quad
  a_j^{\text{AW}}\to-a_j\ (j=1,2),\quad
  a_j^{\text{AW}}\to-a_{j-2}^{*\,-1}\ (j=3,4),
  \label{okikae(vii)}
\end{equation} 
into \eqref{HtPnAW}, we obtain
\begin{align}
  &\quad\frac{\prod_{j=1}^2(1+a_je^x)(1+a_j^{*\,-1}e^x)}
  {(e^{2x}-1)(e^{-i\gamma}e^{2x}-1)}
  \Bigl(p_n\bigl(\cosh(x-i\gamma);-a_1,-a_2,-a_1^{*\,-1},-a_2^{*\,-1}
  |e^{-i\gamma}\bigr)\n
  &\qquad\qquad\qquad\qquad\qquad\qquad\quad
  -p_n\bigl(\cosh x;-a_1,-a_2,-a_1^{*\,-1},-a_2^{*\,-1}|e^{-i\gamma})\Bigr)\n
  &+e^{i\gamma}e^{-2i\gamma\alpha}
  \frac{\prod_{j=1}^2(1+a_j^*e^x)(1+a_j^{-1}e^x)}
  {(e^{2x}-1)(e^{i\gamma}e^{2x}-1)}
  \Bigl(p_n\bigl(\cosh(x+i\gamma);-a_1,-a_2,-a_1^{*\,-1},-a_2^{*\,-1}
  |e^{-i\gamma}\bigr)\n
  &\qquad\qquad\qquad\qquad\qquad\qquad\qquad\qquad
  -p_n\bigl(\cosh x;-a_1,-a_2,-a_1^{*\,-1},-a_2^{*\,-1}|e^{-i\gamma})\Bigr)
  \label{HtPn(vii)}\\
  &=-e^{i\frac{\gamma}{2}}e^{-i\gamma\alpha}\times
  4\sin\tfrac{\gamma}{2}n\,\sin\tfrac{\gamma}{2}(n-1+2\alpha)\,
  p_n\bigl(\cosh x;-a_1,-a_2,-a_1^{*\,-1},-a_2^{*\,-1}|e^{-i\gamma}).
  \nonumber
\end{align}
Dividing this by $-e^{i\frac{\gamma}{2}}e^{-i\gamma\alpha}$ and
multiplying $e^{-i\frac{\pi}{2}n}e^{i\gamma\frac34n(n-1)}e^{i\gamma\alpha n}$
give \eqref{HtPn}.

\subsubsection{groundstate}
\label{sec:coshx,phi0}

To obtain a quantum system $\mathcal{H}$ from a polynomial system
$\widetilde{\mathcal{H}}$, we have to find the groundstate wavefunction
which satisfies \eqref{Aphi0=0} and ensures the hermiticity of the Hamiltonian.
First let us consider the zero mode equation of $\mathcal{A}$, \eqref{Aphi0=0}.
{}From the form of $V(x)$ \eqref{V(vii)}, its main building block is $1+ae^x$
and we want to have a function $f(x)$ satisfying
\begin{equation*}
  \sqrt{1+a^*e^{x-i\frac{\gamma}{2}}}\,f(x-i\tfrac{\gamma}{2})
  =\sqrt{1+ae^{x+i\frac{\gamma}{2}}}\,f(x+i\tfrac{\gamma}{2})
  \quad\text{or}\quad
  \frac{f(x+i\frac{\gamma}{2})^2}{f(x-i\frac{\gamma}{2})^2}
  =\frac{1+e^{x-i\frac{\gamma}{2}+\log a^*}}{1+e^{x+i\frac{\gamma}{2}+\log a}}.
\end{equation*}
The quantum dilogarithm function $\Phi_{\gamma}(z)$ satisfies the functional
equation (see Appendix \ref{app:qdilog})
\begin{equation}
  \frac{\Phi_{\gamma}(z+i\gamma)}{\Phi_{\gamma}(z-i\gamma)}=\frac{1}{1+e^z}.
\end{equation}
{}From this we have
\begin{equation*}
  \frac{\Phi_{\frac{\gamma}{2}}(x+i\frac{\gamma}{2}+\log a+i\frac{\gamma}{2})}
  {\Phi_{\frac{\gamma}{2}}(x+i\frac{\gamma}{2}+\log a-i\frac{\gamma}{2})}
  =\frac{1}{1+e^{x+i\frac{\gamma}{2}+\log a}},\quad
  \frac{\Phi_{\frac{\gamma}{2}}(x-i\frac{\gamma}{2}+\log a^*+i\frac{\gamma}{2})}
  {\Phi_{\frac{\gamma}{2}}(x-i\frac{\gamma}{2}+\log a^*-i\frac{\gamma}{2})}
  =\frac{1}{1+e^{x-i\frac{\gamma}{2}+\log a^*}}.
\end{equation*}
The following $f(x)$,
\begin{equation*}
  f(x)=\biggl(\frac{\Phi_{\frac{\gamma}{2}}(x+i\frac{\gamma}{2}+\log a)}
  {\Phi_{\frac{\gamma}{2}}(x-i\frac{\gamma}{2}+\log a^*)}\biggr)^{\frac12}
  =f^*(x),
\end{equation*}
solves the above equation.
Using this building block, we adopt the groundstate wavefunction
$\phi_0(x;\bm{\lambda})$ as
\begin{align}
  \phi_0(x;\bm{\lambda})
  &=e^{(\frac12-\alpha-\frac{\pi}{\gamma})x}
  \sqrt{(e^{2x}-1)(e^{\frac{4\pi}{\gamma}x}-1)}\n
  &\quad\times\biggl(\prod_{j=1}^2\frac{
  \Phi_{\frac{\gamma}{2}}\bigl(x+\gamma\beta_j+i\gamma(\frac12-\alpha_j)\bigr)
  \Phi_{\frac{\gamma}{2}}\bigl(x-\gamma\beta_j+i\gamma(\frac12-\alpha_j)\bigr)}
  {\Phi_{\frac{\gamma}{2}}\bigl(x+\gamma\beta_j-i\gamma(\frac12-\alpha_j)\bigr)
  \Phi_{\frac{\gamma}{2}}\bigl(x-\gamma\beta_j-i\gamma(\frac12-\alpha_j)\bigr)}
  \biggr)^{\frac12}.
  \label{phi0(vii)}
\end{align}
By using the properties of $\Phi_{\gamma}(z)$ presented in Appendix
\ref{app:qdilog}, we can show that \eqref{Aphi0=0} and
\begin{align}
  \phi_0^*(x;\bm{\lambda})&=\phi_0(x;\bm{\lambda}),
  \label{phi0*}\\
  \phi_0(-x;\bm{\lambda})^2&=\phi_0(x;\bm{\lambda})^2.
  \label{phi0(-x)(vii)}
\end{align}
Whether this groundstate wavefunction ensures the hermiticity of the
Hamiltonian will be clarified in the next \S\,\ref{sec:coshx,hermite}.

As shown in the previous subsection the eigenpolynomials of (\romannumeral7)
and the Askey-Wilson systems (\romannumeral3) are related to each other under
the replacement \eqref{okikae(vii)}.
As for the groundstate wavefunctions the correspondence is rather formal.
After rewriting $\Phi_{\gamma}$ in $\phi_0$ \eqref{phi0(vii)} as
$\Phi_{\gamma}(z)=\Phi^{\text{F}}_{\!\!\!\sqrt{\frac{\pi}{\gamma}}}
\bigl(\frac{z}{2\sqrt{\pi\gamma}}\bigr)^{-1}$ \cite{qdilog3} (see
\eqref{Fqdilog}), we perform the inverse of the replacement
\eqref{okikae(vii)} and use \eqref{Fqdilogprod} to obtain
\begin{align}
  &\phi_0(x;\bm{\lambda})
  \to\phi_0^{\text{AW}}(x^{\text{AW}};\bm{\lambda}^{\text{AW}}
  ;\gamma^{\text{AW}})
  \,\mathcal{C}(x^{\text{AW}},\bm{\lambda}^{\text{AW}},\gamma^{\text{AW}}),\n
  &\phi_0^{\text{AW}}(x;\bm{\lambda};\gamma)
  =\biggl(\frac{(e^{2ix},e^{-2ix};e^{\gamma})_{\infty}}
  {\prod_{j=1}^4(a_je^{ix},a_je^{-ix};e^{\gamma})_{\infty}}
  \biggr)^{\frac12}\ \ (a_j=e^{\gamma\lambda_j}),\\
  &\mathcal{C}(x,\bm{\lambda},\gamma)
  =e^{\frac12(1-\sum_{j=1}^4\lambda_j)ix+\frac{\pi}{\gamma}x}
  \sqrt{(e^{2ix}-1)(e^{\frac{4\pi}{\gamma}x}-1)}\n
  &\qquad\qquad\quad\times
  \biggl(\frac{\prod_{j=1}^4(e^{ix-\gamma\lambda_j+\gamma},
  e^{-ix+\gamma\lambda_j};e^{\gamma})_{\infty}}
  {(e^{2ix},e^{-2ix};e^{\gamma})_{\infty}}
  \prod_{j=1}^4\frac{
  (e^{\frac{2\pi}{\gamma}x-2\pi i\lambda_j+\frac{4\pi^2}{\gamma}};
  e^{\frac{4\pi^2}{\gamma}})_{\infty}}
  {(e^{\frac{2\pi}{\gamma}x+2\pi i\lambda_j};
  e^{\frac{4\pi^2}{\gamma}})_{\infty}}
  \biggr)^{\frac12}.\nonumber
\end{align}
Here $\phi_0^{\text{AW}}$ is the groundstate wavefunction of the
Askey-Wilson system and $\mathcal{C}$ is a pseudo constant,
$\mathcal{C}(x-i\gamma,\bm{\lambda},\gamma)
=\mathcal{C}(x,\bm{\lambda},\gamma)$.
Therefore $\mathcal{C}$ does not contribute to \eqref{Aphi0=0}
and we can discard this.

\subsubsection{hermiticity}
\label{sec:coshx,hermite}

We restrict the range of the parameters as
\begin{equation}
  -\gamma\alpha>\pi+\frac{\gamma}{2},\quad
  \gamma-\pi<\gamma\alpha_j<0\ \ (j=1,2).
  \label{range(vii)}
\end{equation}
The eigenpolynomial $P_n(\eta;\bm{\lambda})$ is expressed in terms of a
terminating ${}_4\phi_3$, and the denominator of each term of the expansion
\eqref{defqhypergeom} is
$(a_1a_2,a_1a_1^{*\,-1},a_1a_2^{*\,-1};e^{-i\gamma})_k\cdot
(e^{-i\gamma};e^{-i\gamma})_k$.
We avoid the parameter values of $\{a_j\}$ such that
$(a_1a_2,a_1a_1^{*\,-1},a_1a_2^{*\,-1};e^{-i\gamma})_k$ vanishes.
Concerning the factor $(e^{-i\gamma};e^{-i\gamma})_k$,
see \S\,\ref{sec:coshx,comment}.

For $x=R+iy$ ($R>0$, $0\leq y\leq\gamma$), the argument $z$ of each
$\Phi_{\frac{\gamma}{2}}(z)$ in $\phi_0(x;\bm{\lambda})$ satisfies
$|\text{Im}\,z|<\frac{\gamma}{2}+\pi$, where the integral representation
\eqref{intrep} and the asymptotic forms \eqref{asymp} are valid.
{}From \eqref{asymp}, the asymptotic behaviour of $\phi_0(R+iy;\bm{\lambda})$
($0\leq y\leq\gamma$) at large $R$ is
\begin{align}
  \bigl|\phi_0(R+iy;\bm{\lambda})\bigr|\simeq\text{const}\times
  e^{(-\frac12+\alpha+\frac{\pi}{\gamma})R},
  \label{phi0(R)(vii)}
\end{align}
and those of $V(R+iy;\bm{\lambda})$ and
$P_n\bigl(\eta(R+iy);\bm{\lambda}\bigr)$ are
\begin{align}
  \bigl|V(R+iy;\bm{\lambda})\bigr|\simeq\text{const},\quad
  \bigl|P_n\bigl(\eta(R+iy);\bm{\lambda}\bigr)\bigr|
  \simeq\text{const}\times e^{nR}.
  \label{VPn(R)(vii)}
\end{align}
At $x=x_1=0$, $\phi_0(x;\bm{\lambda})^2$  and
$P_n\bigl(\eta(x);\bm{\lambda}\bigr)$ are regular.
Therefore the wavefunction $\phi_n(x;\bm{\lambda})=\phi_0(x;\bm{\lambda})
P_n\bigl(\eta(x);\bm{\lambda}\bigr)=\phi_n^*(x;\bm{\lambda})$ is
square integrable $(\phi_n,\phi_n)<\infty$ only for
\begin{equation}
  n<\frac12-\alpha-\frac{\pi}{\gamma}.
  \label{n(vii)}
\end{equation}
The maximal value of $n$ is
$n_{\text{max}}(\bm{\lambda})=[\frac12-\alpha-\frac{\pi}{\gamma}]'$, where
$[x]'$ denotes the greatest integer not exceeding and not equal to $x$.

Let us check the condition \eqref{H=Hd}.
We can show that
\begin{align*}
  \text{(a)}:&\quad
  \frac{n_1+n_2}{2}<\frac12-\alpha-\frac{\pi}{\gamma}\ \ \Rightarrow
  \ \int_{-\frac{\gamma}{2}}^{\frac{\gamma}{2}}dy\,G(x_2+iy)
  =0=\int_{-\frac{\gamma}{2}}^{\frac{\gamma}{2}}dy\,G^*(x_2-iy),\\
  \text{(b)}:&\quad
  \int_{-\frac{\gamma}{2}}^{\frac{\gamma}{2}}dy\,G(x_1+iy)
  =\int_{-\frac{\gamma}{2}}^{\frac{\gamma}{2}}dy\,G^*(x_1-iy),\\
  \text{(c)}:&\quad
  \text{$G(x)$ and $G^*(x)$ do not have poles in $D_{\gamma}$}.
\end{align*}
Here we have used \eqref{phi0(R)(vii)}--\eqref{VPn(R)(vii)} for (a),
\eqref{V*(-x)(vii)} and \eqref{phi0(-x)(vii)} for (b) and
\eqref{pole}--\eqref{zero} for (c).
Thus the condition \eqref{H=Hd} holds for the eigenstates
$\phi_n(x;\bm{\lambda})$ ($n=0,1,\ldots,n_{\text{max}}$),
namely the hermiticity of the Hamiltonian is established.
It is straightforward to verify that the energy eigenvalues are monotonously
increasing: $0=\mathcal{E}_0(\bm{\lambda})<\mathcal{E}_1(\bm{\lambda})
<\cdots<\mathcal{E}_{n_{\text{max}}}(\bm{\lambda})$.

\subsubsection{comments on $\bm{\frac{\gamma}{2\pi}\in\mathbb{Q}}$}
\label{sec:coshx,comment}

Let us set $q=e^{-i\gamma}$.
If $\frac{\gamma}{2\pi}\in\mathbb{Q}$, there are positive integers $m$
leading to $q^m=1$.
The expansion formula of $P_n(\eta;\bm{\lambda})$ has factors
$\frac{1}{(q;q)_k}$ ($1\leq k\leq n$) (see the sentences below
\eqref{range(vii)}), namely factors $\frac{1}{1-q^k}$.
So, if $q^k=1$ happens for some $1\leq k\leq n$, $P_n(\eta;\bm{\lambda})$
diverges and it loses its meaning.
However, this does not happen for the eigenfunctions
$\phi_n(x;\bm{\lambda})$ ($n=0,1,\ldots,n_{\text{max}}$).

Let us set $\gamma=\frac{M}{N}2\pi$ ($N$ and $M$ are coprime positive integers).
The range $0<\gamma<\pi$ means $2M<N$.
The smallest positive integer $m$ leading to  $q^m=1$ is $m=N$.
The ranges of $n$ \eqref{n(vii)} and the parameters \eqref{a_j} mean
$n<\frac12-\alpha-\frac{\pi}{\gamma}<\frac12+\frac{\pi}{\gamma}$.
These lead to  $n<\frac12+\frac{\pi}{\gamma}=\frac12+\frac{N}{2M}
\leq\frac12+\frac{N}{2}<N$.
Therefore  $q^k=1$ ($1\leq k\leq n$) does not happen for
$n\leq n_{\text{max}}$.

\subsection{$\bm{\eta(x)=e^{\pm x}}$}
\label{sec:e^x}

Here we consider the cases (\romannumeral5) $\eta(x)=e^{-x}$ and
(\romannumeral6) $\eta(x)=e^x$, $-\infty<x<\infty$, $0<\gamma<\pi$.
The outline is the same as that in \S\,\ref{sec:coshx} and we will present
the results briefly.

The factorised potential functions \eqref{Vtfactor} derived from the original
parameters form \eqref{exV} read as follows
($A=\frac{4\sin\frac{\gamma}{2}\sin\gamma}{|a_1a_2|}$ in \eqref{Vtfactor}):
\begin{align}
  \text{(\romannumeral5)}:&\quad
  V(x;\bm{\lambda})
  =e^{i\pi}e^{-i\frac{\gamma}{2}}\frac{a_1^*a_2^*}{|a_1a_2|}
  \prod_{j=1}^2(1+a_j^{*\,-1}e^x),
  \label{V(v)}\\[-2pt]
  \text{(\romannumeral6)}:&\quad
  V(x;\bm{\lambda})
  =e^{-i\pi}e^{i\frac{\gamma}{2}}\frac{a_1a_2}{|a_1a_2|}
  \prod_{j=1}^2(1+a_j^{-1}e^{-x}),
  \label{V(vi)}
\end{align}
where $e^{\pm i\pi}$ indicates $\sqrt{e^{\pm i\pi}}=e^{\pm\frac{i}{2}\pi}$
in $\sqrt{V(x;\bm{\lambda})}$, and
the relation between $\bm{\lambda}=(\lambda_1,\lambda_2)$ and
$a_j\in\mathbb{C}_{\neq 0}$ ($j=1,2$) is \eqref{a_j}.
Note that these two systems (\romannumeral5) and (\romannumeral6) are
essentially the same, as they are related by the replacements
$x\leftrightarrow-x$ and $V(x)\leftrightarrow V^*(-x)$:
\begin{equation}
  V^{\text{(\romannumeral6)}}(x;\bm{\lambda})
  =V^{\text{(\romannumeral5)\,*}}(-x;\bm{\lambda}).
  \label{VV*-}
\end{equation}

The energy eigenvalue $\mathcal{E}_n(\bm{\lambda})$ and the corresponding
eigenpolynomial $P_n(\eta;\bm{\lambda})$ are listed in \eqref{exEn} and
\eqref{exPndet}:
\begin{align}
  \mathcal{E}_n(\bm{\lambda})&
  =4\sin\tfrac{\gamma}{2}n\,\sin\tfrac{\gamma}{2}(n-1+2\alpha),
  \label{En(v)}\\
  P_n(\eta;\bm{\lambda})&
  =e^{-i\frac{\pi}{2}n}e^{i\gamma\frac34n(n-1)}e^{i\gamma\alpha n}\,
  \tilde{p}_n(\eta;-a_1,-a_2,-a_1^{*\,-1},-a_2^{*\,-1}|e^{-i\gamma}),
  \label{Pn(v)}
\end{align}
where $\tilde{p}_n$ is defined in \eqref{ptn} as obtained from the
Askey-Wilson polynomial by a certain limiting procedure.
By using \eqref{ptn} and \eqref{ptnprop1}--\eqref{ptnprop2}, we can show
the `reality' of $P_n$ \eqref{Pn*}.
This eigenpolynomial \eqref{Pn(v)} can be obtained from that of the case
(\romannumeral7) \eqref{Pn(vii)} (with an appropriate overall normalisation)
by replacement ($R\in\mathbb{R}$)
\begin{alignat}{2}
  \text{(\romannumeral5)}:&\quad
  x^{\text{(\romannumeral7)}}=-x+R,&\quad
  &a_j^{\text{(\romannumeral7)}}=a_j^{*\,-1}e^R\ (j=1,2),
  \label{okikae(v)}\\
  \text{(\romannumeral6)}:&\quad
  x^{\text{(\romannumeral7)}}=x+R,&\quad
  &a_j^{\text{(\romannumeral7)}}=a_je^{-R}\ (j=1,2),
  \label{okikae(vi)}
\end{alignat}
and the limit $R\to\infty$.
The above factorised potential functions \eqref{V(v)}--\eqref{V(vi)} and the
polynomial equation \eqref{HtPn} can also be obtained from those of the case
(\romannumeral7) \eqref{V(vii)} and \eqref{HtPn(vii)} by the same limit.

We adopt the following groundstate wavefunctions $\phi_0(x;\bm{\lambda})$:
\begin{align}
  \text{(\romannumeral5)}:&\quad
  \phi_0(x;\bm{\lambda})
  =e^{(\frac12-\alpha-\frac{\pi}{\gamma})x}
  \biggl(\prod_{j=1}^2\frac{
  \Phi_{\frac{\gamma}{2}}\bigl(x-\gamma\beta_j+i\gamma(\frac12-\alpha_j)\bigr)}
  {\Phi_{\frac{\gamma}{2}}\bigl(x-\gamma\beta_j-i\gamma(\frac12-\alpha_j)\bigr)}
  \biggr)^{\frac12},
  \label{phi0(v)}\\
  \text{(\romannumeral6)}:&\quad
  \phi_0(x;\bm{\lambda})
  =e^{-(\frac12-\alpha-\frac{\pi}{\gamma})x}
  \biggl(\prod_{j=1}^2\frac{
  \Phi_{\frac{\gamma}{2}}\bigl(-x-\gamma\beta_j+i\gamma(\frac12-\alpha_j)\bigr)}
  {\Phi_{\frac{\gamma}{2}}\bigl(-x-\gamma\beta_j-i\gamma(\frac12-\alpha_j)
  \bigr)}
  \biggr)^{\frac12}.
  \label{phi0(vi)}
\end{align}
Reflecting the relationship of the potentials \eqref{VV*-} these two
groundstate wavefunctions are also related:
\begin{equation}
  \phi_0^{\text{(\romannumeral6)}}(x;\bm{\lambda})
  =\phi_0^{\text{(\romannumeral5)}}(-x;\bm{\lambda}).
  \label{phi0(-x)(v)}
\end{equation}
The zero mode equation \eqref{Aphi0=0} and the `reality' of $\phi_0$
\eqref{phi0*} can be derived from the properties of $\Phi_{\gamma}(z)$
presented in Appendix \ref{app:qdilog}.
These groundstate wavefunctions \eqref{phi0(v)}--\eqref{phi0(vi)} can be
obtained from that of the case (\romannumeral7) \eqref{phi0(vii)} (with an
appropriate overall normalisation) by the replacement
\eqref{okikae(v)}--\eqref{okikae(vi)} and $R\to\infty$ limit.

We restrict the range of the parameters as
\begin{equation}
  -\gamma\alpha>\pi+\frac{\gamma}{2},\quad
  \gamma-\pi<\gamma\alpha_j<0\ \ (j=1,2).
  \label{range(v)}
\end{equation}
We avoid the parameter values of $\{a_j\}$ such that the denominators of
$P_n$ vanish.
Since (\romannumeral5) and (\romannumeral6) are essentially the same,
we will argue the case (\romannumeral5) below.
For $x=\pm R+iy$ ($R>0$, $0\leq y\leq\gamma$), the asymptotic behaviour of
$\phi_0(\pm R+iy;\bm{\lambda})$ at large $R$ is
\begin{align}
  \bigl|\phi_0(R+iy;\bm{\lambda})\bigr|\simeq\text{const}\times
  e^{(-\frac12-\frac{\pi}{\gamma})R},\quad
  \bigl|\phi_0(-R+iy;\bm{\lambda})\bigr|\simeq\text{const}\times
  e^{-(\frac12-\alpha-\frac{\pi}{\gamma})R},
  \label{phi0(R)(v)}
\end{align}
and those of $V(\pm R+iy;\bm{\lambda})$ and
$P_n\bigl(\eta(\pm R+iy);\bm{\lambda}\bigr)$ are
\begin{alignat}{2}
  \bigl|V(R+iy;\bm{\lambda})\bigr|&\simeq\text{const}\times e^{2R},&
  \bigl|V(-R+iy;\bm{\lambda})\bigr|&\simeq\text{const},\n
  \bigl|P_n\bigl(\eta(R+iy);\bm{\lambda}\bigr)\bigr|
  &\simeq\text{const},&
  \bigl|P_n\bigl(\eta(-R+iy);\bm{\lambda}\bigr)\bigr|
  &\simeq\text{const}\times e^{nR}.
  \label{VPn(R)(v)}
\end{alignat}
The wavefunction $\phi_n(x;\bm{\lambda})=\phi_0(x;\bm{\lambda})
P_n\bigl(\eta(x);\bm{\lambda}\bigr)=\phi_n^*(x;\bm{\lambda})$ is
square integrable $(\phi_n,\phi_n)<\infty$ only for
\begin{equation}
  n<\frac12-\alpha-\frac{\pi}{\gamma}.
  \label{n(v)}
\end{equation}
The maximal value of $n$ is
$n_{\text{max}}(\bm{\lambda})=[\frac12-\alpha-\frac{\pi}{\gamma}]'$.
We can show that
\begin{align*}
  \text{(a)}:&\quad
  \int_{-\frac{\gamma}{2}}^{\frac{\gamma}{2}}dy\,G(x_2+iy)
  =0=\int_{-\frac{\gamma}{2}}^{\frac{\gamma}{2}}dy\,G^*(x_2-iy),\\
  \text{(b)}:&\quad
  \frac{n_1+n_2}{2}<\frac12-\alpha-\frac{\pi}{\gamma}\ \ \Rightarrow
  \ \int_{-\frac{\gamma}{2}}^{\frac{\gamma}{2}}dy\,G(x_1+iy)
  =0=\int_{-\frac{\gamma}{2}}^{\frac{\gamma}{2}}dy\,G^*(x_1-iy),\\
  \text{(c)}:&\quad
  \text{$G(x)$ and $G^*(x)$ do not have poles in $D_{\gamma}$}.
\end{align*}
Here \eqref{phi0(R)(v)}--\eqref{VPn(R)(v)} are used for (a) and (b),
and \eqref{pole}--\eqref{zero} for (c).
Thus the condition \eqref{H=Hd} holds for the eigenstates
$\phi_n(x;\bm{\lambda})$ ($n=0,1,\ldots,n_{\text{max}}$),
namely the hermiticity of the Hamiltonian is established.
We can verify that
$0=\mathcal{E}_0(\bm{\lambda})<\mathcal{E}_1(\bm{\lambda})
<\cdots<\mathcal{E}_{n_{\text{max}}}(\bm{\lambda})$.

As in the $\cosh x$ example \S\,\ref{sec:coshx,comment}, the `rational'
$\frac{\gamma}{2\pi}\in\mathbb{Q}$ does not cause any trouble for the
eigenfunctions $\phi_n(x;\bm{\lambda})$ ($n=0,1,\ldots,n_{\text{max}}$).

\subsection{$\bm{\eta(x)=\sinh x}$}
\label{sec:sinhx}

Here we consider the case (\romannumeral8) $\eta(x)=\sinh x$,
$-\infty<x<\infty$, $0<\gamma<\pi$.
The outline is the same as that in \S\,\ref{sec:coshx} and we will present
the results briefly.

The factorised potential functions \eqref{Vtfactor} derived from the original
parameters form \eqref{exV} read
($A=(-1)^{K+1}\frac{\sin\frac{\gamma}{2}\sin\gamma}{|a_1a_2|}$ in
\eqref{Vtfactor}):
\begin{equation}
  V(x;\bm{\lambda})
  =e^{i\pi K}e^{-i\frac{\gamma}{2}}\frac{a_1^*a_2^*}{|a_1a_2|}
  \frac{\prod_{j=1}^2(1+a_je^x)(1-a_j^{*\,-1}e^x)}
  {(1+e^{2x})(1+e^{-i\gamma}e^{2x})}\quad(K=\pm1,0),
  \label{V(viii)}
\end{equation}
where $e^{i\pi K}$ indicates $\sqrt{e^{i\pi K}}=e^{\frac{i}{2}\pi K}$
in $\sqrt{V(x;\bm{\lambda})}$, and
the relation between $\bm{\lambda}=(\lambda_1,\lambda_2)$ and
$a_j\in\mathbb{C}_{\neq 0}$ ($j=1,2$) is \eqref{a_j}.
The three choices of $K$ correspond to three different parameter ranges given
below, see \eqref{range(viii)}.

The energy eigenvalue $\mathcal{E}_n(\bm{\lambda})$ and the corresponding
eigenpolynomial  $P_n(\eta;\bm{\lambda})$ are listed in \eqref{exEn}
and  \eqref{exPndet}:
\begin{align}
  \mathcal{E}_n(\bm{\lambda})&
  =(-1)^{K+1}4\sin\tfrac{\gamma}{2}n\,\sin\tfrac{\gamma}{2}(n-1+2\alpha),
  \label{En(viii)}\\
  P_n(\eta;\bm{\lambda})&
  =e^{-i\frac{\pi}{2}n}e^{i\gamma\frac34n(n-1)}e^{i\gamma\alpha n}\,
  (-i)^np_n(i\eta;ia_1,ia_2,-ia_1^{*\,-1},-ia_2^{*\,-1}|e^{-i\gamma}),
  \label{Pn(viii)}
\end{align}
where $p_n$ is the Askey-Wilson polynomial \eqref{AWpoly}.
The `reality' of $P_n$ \eqref{Pn*} can be shown based on the properties
\eqref{pnprop1}--\eqref{pnprop3} and the expansion formula of the
Askey-Wilson polynomial  \eqref{AWpoly}.
The polynomial equation \eqref{HtPn} can be obtained from that of the
Askey-Wilson case \eqref{HtPnAW} by substitution
\begin{equation}
  x^{\text{AW}}\to-ix+\frac{\pi}{2},\quad
  \gamma^{\text{AW}}\to-i\gamma,\quad
  a_j^{\text{AW}}\to ia_j\ (j=1,2),\quad
  a_j^{\text{AW}}\to-ia_{j-2}^{*\,-1}\ (j=3,4).
  \label{okikae(viii)}
\end{equation}

We adopt the following groundstate wavefunction $\phi_0(x;\bm{\lambda})$:
\begin{align}
  \phi_0(x;\bm{\lambda})
  &=e^{(\frac12-\alpha-K\frac{\pi}{\gamma})x}\sqrt{1+e^{2x}}\n
  &\quad\times\biggl(\prod_{j=1}^2\frac{
  \Phi_{\frac{\gamma}{2}}\bigl(x+\gamma\beta_j+i\gamma(\frac12-\alpha_j)\bigr)
  \Phi_{\frac{\gamma}{2}}\bigl(x-\gamma\beta_j+i\gamma(\frac12-\alpha_j^-)
  \bigr)}
  {\Phi_{\frac{\gamma}{2}}\bigl(x+\gamma\beta_j-i\gamma(\frac12-\alpha_j)\bigr)
  \Phi_{\frac{\gamma}{2}}\bigl(x-\gamma\beta_j-i\gamma(\frac12-\alpha_j^-)
  \bigr)}
  \biggr)^{\frac12},
  \label{phi0(viii)}
\end{align}
where $\alpha_j^-$ ($j=1,2$) is defined by
\begin{equation}
  \gamma\alpha_j^-\eqdef\left\{
  \begin{array}{ll}
  \gamma\alpha_j+\pi&\text{for $-\pi<\gamma\alpha_j\leq 0$}\\[2pt]
  \gamma\alpha_j-\pi&\text{for $\ \ \,0<\gamma\alpha_j\leq\pi$}
  \end{array}\right.,\quad
  \alpha^-\eqdef\alpha_1^-+\alpha_2^-,
  \label{alpha-}
\end{equation}
and it satisfies $-\pi<\gamma\alpha_j^-\leq\pi$.
We can verify the zero mode equation \eqref{Aphi0=0} and the `reality' of
$\phi_0$ \eqref{phi0*} based on the properties of the quantum dilogarithm
function $\Phi_{\gamma}(z)$ presented in Appendix \ref{app:qdilog}.

We restrict the range of the parameters as
\begin{equation}
  -\gamma\alpha>K\pi+\frac{\gamma}{2},\quad
  \left\{
  \begin{array}{ll}
  K=1:&\gamma-\pi<\gamma\alpha_j<0\ \ (j=1,2)\\[2pt]
  K=-1:&\gamma<\gamma\alpha_j<\pi\ \ (j=1,2)\\[2pt]
  K=0:&\gamma-\pi<\gamma\alpha_1<0,\ \gamma<\gamma\alpha_2<\pi\\
  &\!\!\!\!\!\!\!\text{or}\ \gamma-\pi<\gamma\alpha_2<0,
  \ \gamma<\gamma\alpha_1<\pi
  \end{array}\right..
  \label{range(viii)}
\end{equation}
We avoid the parameter values of $\{a_j\}$ such that the denominators of
$P_n$ vanish.
For $x=\pm R+iy$ ($R>0$, $0\leq y\leq\gamma$),
the asymptotic behaviour of $\phi_0(\pm R+iy;\bm{\lambda})$ at large $R$ is
\begin{equation}
  \bigl|\phi_0(R+iy;\bm{\lambda})\bigr|\simeq\text{const}\times
  e^{(-\frac12+\alpha^--\frac{K\pi}{\gamma})R},
  \ \ \bigl|\phi_0(-R+iy;\bm{\lambda})\bigr|\simeq\text{const}\times
  e^{-(\frac12-\alpha-\frac{K\pi}{\gamma})R},
  \label{phi0(R)(viii)}
\end{equation}
and those of $V(\pm R+iy;\bm{\lambda})$ and
$P_n\bigl(\eta(\pm R+iy);\bm{\lambda}\bigr)$ are
\begin{equation}
  \bigl|V(\pm R+iy;\bm{\lambda})\bigr|\simeq\text{const},\quad
  \bigl|P_n\bigl(\eta(\pm R+iy);\bm{\lambda}\bigr)\bigr|
  \simeq\text{const}\times e^{nR}.
  \label{VPn(R)(viii)}
\end{equation}
The wavefunction $\phi_n(x;\bm{\lambda})=\phi_0(x;\bm{\lambda})
P_n\bigl(\eta(x);\bm{\lambda}\bigr)=\phi_n^*(x;\bm{\lambda})$ is
square integrable $(\phi_n,\phi_n)<\infty$ only for
\begin{equation}
  n<\frac12-\alpha-\frac{K\pi}{\gamma}.
  \label{n(viii)}
\end{equation}
The maximal value of $n$ is
$n_{\text{max}}(\bm{\lambda})=[\frac12-\alpha-\frac{K\pi}{\gamma}]'$.
We can show that
\begin{align*}
  \text{(a)}:&\quad
  \frac{n_1+n_2}{2}<\frac12-\alpha^-+\frac{K\pi}{\gamma}\ \ \Rightarrow
  \ \int_{-\frac{\gamma}{2}}^{\frac{\gamma}{2}}dy\,G(x_2+iy)
  =0=\int_{-\frac{\gamma}{2}}^{\frac{\gamma}{2}}dy\,G^*(x_2-iy),\\
  \text{(b)}:&\quad
  \frac{n_1+n_2}{2}<\frac12-\alpha-\frac{K\pi}{\gamma}\ \ \Rightarrow
  \ \int_{-\frac{\gamma}{2}}^{\frac{\gamma}{2}}dy\,G(x_1+iy)
  =0=\int_{-\frac{\gamma}{2}}^{\frac{\gamma}{2}}dy\,G^*(x_1-iy),\\
  \text{(c)}:&\quad
  \text{$G(x)$ and $G^*(x)$ do not have poles in $D_{\gamma}$}.
\end{align*}
Here various asymptotic behaviours \eqref{phi0(R)(viii)}--\eqref{VPn(R)(viii)}
are used for (a) and (b), and the positions of the poles and zeros of the
quantum dilogarithm \eqref{pole}--\eqref{zero} for (c).
Thus the hermiticity condition \eqref{H=Hd} holds for the eigenstates
$\phi_n(x;\bm{\lambda})$ ($n=0,1,\ldots,n_{\text{max}}$), namely the
Hamiltonian is hermitian.
We can verify that
$0=\mathcal{E}_0(\bm{\lambda})<\mathcal{E}_1(\bm{\lambda})
<\cdots<\mathcal{E}_{n_{\text{max}}}(\bm{\lambda})$.

As in  \S\,\ref{sec:coshx,comment}, the case
$\frac{\gamma}{2\pi}\in\mathbb{Q}$ does not cause any trouble for the
eigenstates $\phi_n(x;\bm{\lambda})$ ($n=0,1,\ldots,n_{\text{max}}$),
because we have
$n<\frac12-\alpha-\frac{K\pi}{\gamma}<\frac12+\frac{\pi}{\gamma}$ due to
\eqref{n(viii)}.

\subsection{Closure relation}
\label{sec:cr}

The closure relation is a sufficient condition for exact solvability of
quantum systems whose eigenfunctions have the factorised form with the
sinusoidal coordinate \eqref{phin_form}.
It is a commutator relation between the Hamiltonian $\mathcal{H}$
(or $\widetilde{\mathcal{H}}$) and the sinusoidal coordinate $\eta(x)$
\cite{os7}:
\begin{align}
  &[\mathcal{H},[\mathcal{H},\eta]]=\eta R_0(\mathcal{H})
  +[\mathcal{H},\eta]R_1(\mathcal{H})+R_{-1}(\mathcal{H}),
  \label{cr}\\
  \text{or}\quad&
  [\widetilde{\mathcal{H}},[\widetilde{\mathcal{H}},\eta]]
  =\eta R_0(\widetilde{\mathcal{H}})
  +[\widetilde{\mathcal{H}},\eta]R_1(\widetilde{\mathcal{H}})
  +R_{-1}(\widetilde{\mathcal{H}}),
  \label{crt}
\end{align}
where $R_i(z)$ are polynomials in $z$ with real coefficients $r_i^{(j)}$,
\begin{equation}
  R_1(z)=r_1^{(1)}z+r_1^{(0)},\quad
  R_0(z)=r_0^{(2)}z^2+r_0^{(1)}z+r_0^{(0)},\quad
  R_{-1}(z)=r_{-1}^{(2)}z^2+r_{-1}^{(1)}z+r_{-1}^{(0)}.
  \label{Ricoeff}
\end{equation}
The constants $r_1^{(1)}$ and $r_{-1}^{(2)}$ have appeared in \eqref{eta_os14}.
The closure relation \eqref{cr} allows us to obtain the exact Heisenberg
operator solution for $\eta(x)$, and the annihilation and creation operators
$a^{(\pm)}$ are extracted from this exact Heisenberg operator solution
\cite{os7}.
Roughly speaking, the three terms in r.h.s of \eqref{cr} correspond to the
three term recurrence relations of the corresponding orthogonal polynomials.

Exactly solvable Hamiltonians presented in \S\,\ref{sec:os14} satisfy the
closure relation \eqref{cr}. The explicit expressions of $r_i^{(j)}$ in
terms of $v_{k,l}$ are given in (\U 3.16)--(\U 3.18).
For the new examples (\romannumeral5)--(\romannumeral8), the coefficients
$r_i^{(j)}$ in \eqref{Ricoeff} in terms of the factorisation parameters are
\begin{align}
  &r_{-1}^{(2)}=0,\quad
  r_0^{(2)}=r_1^{(1)},\quad
  r_0^{(1)}=2r_1^{(0)},\quad
  r_1^{(1)}=(e^{i\frac{\gamma}{2}}-e^{-i\frac{\gamma}{2}})^2,\n
  &r_1^{(0)}=-(e^{i\frac{\gamma}{2}}-e^{-i\frac{\gamma}{2}})^2
  (e^{-i\frac{\gamma}{2}}e^{i\gamma\alpha}
  +e^{i\frac{\gamma}{2}}e^{-i\gamma\alpha})\times C,\n
  &r_0^{(0)}=(e^{i\frac{\gamma}{2}}-e^{-i\frac{\gamma}{2}})^2
  (e^{i\gamma\alpha}-e^{-i\gamma\alpha})
  (e^{-i\gamma}e^{i\gamma\alpha}-e^{i\gamma}e^{-i\gamma\alpha}),\\
  &r_{-1}^{(1)}=-(e^{i\frac{\gamma}{2}}-e^{-i\frac{\gamma}{2}})^2
  \bigl((e^{-i\frac{\gamma}{2}}e^{i\gamma\alpha_1}
  +e^{i\frac{\gamma}{2}}e^{-i\gamma\alpha_1})B_2
  +(e^{-i\frac{\gamma}{2}}e^{i\gamma\alpha_2}
  +e^{i\frac{\gamma}{2}}e^{-i\gamma\alpha_2})B_1\bigr)\times C,\n
  &r_{-1}^{(0)}=(e^{i\frac{\gamma}{2}}-e^{-i\frac{\gamma}{2}})^2
  (e^{-i\gamma}e^{i\gamma\alpha}-e^{i\gamma}e^{-i\gamma\alpha})
  \bigl((e^{i\gamma\alpha_1}-e^{-i\gamma\alpha_1})B_2
  +(e^{i\gamma\alpha_2}-e^{-i\gamma\alpha_2})B_1\bigr),\nonumber
\end{align}
where $B_j$ and $C$ are given by
\begin{equation}
  B_j=\left\{
  \begin{array}{ll}
  e^{-\gamma\beta_j}&:\text{(\romannumeral5),\,(\romannumeral6)}\\[2pt]
  \frac12(e^{-\gamma\beta_j}+e^{\gamma\beta_j})&:\text{(\romannumeral7)}\\[4pt]
  \frac12(e^{-\gamma\beta_j}-e^{\gamma\beta_j})&:\text{(\romannumeral8)}
  \end{array}\right.,\quad
  C=\left\{
  \begin{array}{ll}
  1&:\text{(\romannumeral5),\,(\romannumeral6),\,(\romannumeral7)}\\
  (-1)^{K+1}&:\text{(\romannumeral8)}
  \end{array}\right..
\end{equation}

\subsection{Shape invariance}
\label{sec:si}

The shape invariance \cite{genden} is also a sufficient condition for exact
solvability of quantum systems, but its eigenfunctions are not restricted
to the form \eqref{phin_form}.
The shape invariance condition is \cite{os13}
\begin{align}
  &\mathcal{A}(\bm{\lambda})\mathcal{A}(\bm{\lambda})^{\dagger}
  =\kappa\mathcal{A}(\bm{\lambda'})^{\dagger}
  \mathcal{A}(\bm{\lambda'})+\mathcal{E}_1(\bm{\lambda}),
  \label{shapeinv}\\
  \text{or}\quad&
  \mathcal{A}(\bm{\lambda})\mathcal{A}(\bm{\lambda})^{\dagger}
  =\kappa\mathcal{A}(\bm{\lambda}+\bm{\delta})^{\dagger}
  \mathcal{A}(\bm{\lambda}+\bm{\delta})+\mathcal{E}_1(\bm{\lambda}),
  \label{si}
\end{align}
where $\kappa$ is a real positive parameter and $\bm{\lambda}'$ is
uniquely determined by $\bm{\lambda}$.
In concrete examples, with properly chosen parametrisation $\bm{\lambda}$,
$\bm{\lambda}'$ has a simple additive form
$\bm{\lambda}'=\bm{\lambda}+\bm{\delta}$.
As shown in (\U 3.46)--(\U 3.47),
the energy spectrum and the excited state wavefunctions are determined
by the data of the groundstate wavefunction $\phi_0(x;\bm{\lambda})$
and the energy of the first excited state $\mathcal{E}_1(\bm{\lambda})$.

The exactly solvable Hamiltonians presented in \S\,\ref{sec:os14} satisfy
the shape invariance condition \eqref{shapeinv}, since each of the sinusoidal
coordinate $\eta(x)$ \eqref{etalist_os14} satisfies on top of
\eqref{eta_os14} one more condition
\begin{equation}
  \eta(x)-\eta(0)=[\![\tfrac12]\!]\bigl(
  \eta(x-i\tfrac{\gamma}{2})+\eta(x+i\tfrac{\gamma}{2})
  -\eta(-i\tfrac{\gamma}{2})-\eta(i\tfrac{\gamma}{2})\bigr),
  \label{etahalf}
\end{equation}
where $[\![\frac12]\!]$ is defined in \eqref{[[n]]}.
The explicit formulas of $\bm{\lambda}'$ are given in (\U 3.51)--(\U 3.55).

For the new examples (\romannumeral5)--(\romannumeral8), the data of shape
invariance is
\begin{align}
  \bm{\lambda}=(\lambda_1,\lambda_2)=(\alpha_1+i\beta_1,\alpha_2+i\beta_2),
  \quad\bm{\delta}=(\tfrac12,\tfrac12),\quad\kappa=1.
  \label{delta}
\end{align}
Note that the shift $\bm{\lambda}\to\bm{\lambda}+\bm{\delta}$ corresponds
to $a_j\to a_j q^{\frac12}$ ($q=e^{-i\gamma}$), which has the same form as
the Askey-Wilson case.
Let us introduce the auxiliary function $\varphi(x)$,
\begin{equation}
  \varphi(x)\eqdef\left\{
  \begin{array}{ll}
  e^{-x}&:\text{(\romannumeral5)}\\
  e^x&:\text{(\romannumeral6)}\\
  2\sinh x&:\text{(\romannumeral7)}\\
  2\cosh x&:\text{(\romannumeral8)}
  \end{array}\right..
  \label{varphi}
\end{equation}
Then $V(x;\bm{\lambda})$ and $\phi_0(x;\bm{\lambda})$ satisfy
\begin{align}
  V(x;\bm{\lambda}+\bm{\delta})
  &=\kappa^{-1}\frac{\varphi(x-i\gamma)}{\varphi(x)}
  V(x-i\tfrac{\gamma}{2};\bm{\lambda}),
  \label{Vprop}\\
  \phi_0(x;\bm{\lambda}+\bm{\delta})
  &=\varphi(x)\sqrt{V(x+i\tfrac{\gamma}{2};\bm{\lambda})}\,
  \phi_0(x+i\tfrac{\gamma}{2};\bm{\lambda}).
  \label{phi0prop}
\end{align}
The forward and backward shift operators are defined by
\begin{align}
  \mathcal{F}(\bm{\lambda})&\eqdef
  \phi_0(x;\bm{\lambda}+\bm{\delta})^{-1}\circ\mathcal{A}(\bm{\lambda})\circ
  \phi_0(x;\bm{\lambda})=i\varphi(x)^{-1}
  \bigl(e^{\frac{\gamma}{2}p}-e^{-\frac{\gamma}{2}p}\bigr),
  \label{Fshift}\\
  \mathcal{B}(\bm{\lambda})&\eqdef
  \phi_0(x;\bm{\lambda})^{-1}\circ\mathcal{A}(\bm{\lambda})^{\dagger}\circ
  \phi_0(x;\bm{\lambda}+\bm{\delta})
  =-i\bigl(V(x;\bm{\lambda})e^{\frac{\gamma}{2}p}
  -V^*(x;\bm{\lambda})e^{-\frac{\gamma}{2}p}\bigr)
  \varphi(x),
  \label{Bshift}\\
  \widetilde{\mathcal{H}}(\bm{\lambda})&=
  \mathcal{B}(\bm{\lambda})\mathcal{F}(\bm{\lambda}),
\end{align}
which are square root free.
The actions of the operators $\mathcal{A}(\bm{\lambda})$ and
$\mathcal{A}(\bm{\lambda})^{\dagger}$ on the eigenfunctions are
\begin{align}
  \mathcal{A}(\bm{\lambda})\phi_n(x;\bm{\lambda})
  &=f_n(\bm{\lambda})
  \phi_{n-1}\bigl(x;\bm{\lambda}+\bm{\delta}\bigr),
  \label{Aphi=}\\
  \mathcal{A}(\bm{\lambda})^{\dagger}
  \phi_{n-1}\bigl(x;\bm{\lambda}+\bm{\delta}\bigr)
  &=b_{n-1}(\bm{\lambda})\phi_n(x;\bm{\lambda}),
  \label{Adphi=}
\end{align}
and those of $\mathcal{F}$ and $\mathcal{B}$ on the polynomials are
\begin{align}
  \mathcal{F}(\bm{\lambda})P_n\bigl(\eta(x);\bm{\lambda}\bigr)
  &=f_n(\bm{\lambda})P_{n-1}\bigl(\eta(x);\bm{\lambda}+\bm{\delta}\bigr),
  \label{FPn}\\
  \mathcal{B}(\bm{\lambda})P_{n-1}\bigl(\eta(x);\bm{\lambda}+\bm{\delta}\bigr)
  &=b_{n-1}(\bm{\lambda})P_n\bigl(\eta(x);\bm{\lambda}\bigr),
  \label{BPn}
\end{align}
where the real constants $f_n(\bm{\lambda})$ and $b_{n-1}(\bm{\lambda})$
are the factors of the energy eigenvalue
$\mathcal{E}_n(\bm{\lambda})=f_n(\bm{\lambda})b_{n-1}(\bm{\lambda})$ and
we set $P_{-1}(\eta;\bm{\lambda})\eqdef 0$, $b_{-1}(\bm{\lambda})\eqdef 0$.
For (\romannumeral5)--(\romannumeral8), the coefficients $f_n(\bm{\lambda})$
and $b_{n-1}(\bm{\lambda})$ are
\begin{equation}
  f_n(\bm{\lambda})=\mathcal{E}_n(\bm{\lambda})\times\left\{
  \begin{array}{ll}
  -1&:\text{(\romannumeral5)}\\
  1&:\text{(\romannumeral6),\,(\romannumeral7)}\\
  (-1)^{K+1}&:\text{(\romannumeral8)}
  \end{array}\right.,\quad
  b_{n-1}(\bm{\lambda})=\left\{
  \begin{array}{ll}
  -1&:\text{(\romannumeral5)}\\
  1&:\text{(\romannumeral6),\,(\romannumeral7)}\\
  (-1)^{K+1}&:\text{(\romannumeral8)}
  \end{array}\right..
  \label{fnbn}
\end{equation}
We remark that the relations \eqref{FPn}--\eqref{BPn} for (\romannumeral7)
and (\romannumeral8) are obtained from those for the Askey-Wilson case by the
replacements \eqref{okikae(vii)} and \eqref{okikae(viii)}, and
the cases (\romannumeral5) and (\romannumeral6) are obtained from
(\romannumeral7) by the $R\to\infty$ limit with \eqref{okikae(v)} and
\eqref{okikae(vi)}.

\subsection{Normalisation constants}
\label{sec:hn}

The shape invariance gives a recurrence relation of the normalisation constants
$h_n(\bm{\lambda})$ in \eqref{ortho}.
Since the relations \eqref{Aphi=}--\eqref{Adphi=} imply
\begin{align*}
  &\quad b_{n-1}(\bm{\lambda})h_n(\bm{\lambda})
  =\bigl(b_{n-1}(\bm{\lambda})\phi_n(x;\bm{\lambda}),
  \phi_n(x;\bm{\lambda})\bigr)
  =\bigl(\mathcal{A}(\bm{\lambda})^{\dagger}
  \phi_{n-1}(x;\bm{\lambda}+\bm{\delta}),\phi_n(x;\bm{\lambda})\bigr)\\
  &=\bigl(\phi_{n-1}(x;\bm{\lambda}+\bm{\delta}),
  \mathcal{A}(\bm{\lambda})\phi_n(x;\bm{\lambda})\bigr)
  =\bigl(\phi_{n-1}(x;\bm{\lambda}+\bm{\delta}),
  f_n(\bm{\lambda})\phi_{n-1}(x;\bm{\lambda}+\bm{\delta})\bigr)\\
  &=f_n(\bm{\lambda})h_{n-1}(\bm{\lambda}+\bm{\delta}),
\end{align*}
we obtain
\begin{equation}
  h_n(\bm{\lambda})=\frac{f_n(\bm{\lambda})}{b_{n-1}(\bm{\lambda})}
  h_{n-1}(\bm{\lambda}+\bm{\delta})\quad(1\leq n\leq n_{\text{max}}),
  \label{hn->hn-1}
\end{equation}
namely,
\begin{equation}
  h_n(\bm{\lambda})=\prod_{k=1}^n
  \frac{f_k\bigl(\bm{\lambda}+(n-k)\bm{\delta}\bigr)}
  {b_{k-1}\bigl(\bm{\lambda}+(n-k)\bm{\delta}\bigr)}\cdot
  h_0(\bm{\lambda}+n\bm{\delta})\quad(0\leq n\leq n_{\text{max}}).
  \label{hnprop}
\end{equation}

The three term recurrence relations for the orthogonal polynomials
$P_n(\eta;\bm{\lambda})$ ($\text{deg}\,P_n(\eta)=n$, $n\in\mathbb{Z}_{\geq 0}$)
read:
\begin{equation}
  \eta P_n(\eta;\bm{\lambda})=A_n(\bm{\lambda})P_{n+1}(\eta;\bm{\lambda})
  +B_n(\bm{\lambda})P_n(\eta;\bm{\lambda})
  +C_n(\bm{\lambda})P_{n-1}(\eta;\bm{\lambda}).
  \label{3term}
\end{equation}
Let us set
$P_n(\eta;\bm{\lambda})=c_n(\bm{\lambda})P_n^{\text{monic}}(\eta;\bm{\lambda})
=c_n(\bm{\lambda})\eta^n+(\text{lower order terms})$, which gives the
relation $A_n(\bm{\lambda})=\frac{c_n(\bm{\lambda})}{c_{n+1}(\bm{\lambda})}$
because $\eta P^{\text{monic}}_n(\eta;\bm{\lambda})
=P^{\text{monic}}_{n+1}(\eta;\bm{\lambda})+\cdots$.
The explicit forms of $A_n(\bm{\lambda})$, $B_n(\bm{\lambda})$,
$C_n(\bm{\lambda})$ and $c_n(\bm{\lambda})$ for $P_n(\eta;\bm{\lambda})$
can be read from \eqref{pncn}--\eqref{ptn3term} and their definitions
\eqref{Pn(vii)}, \eqref{Pn(v)} and \eqref{Pn(viii)}.
The three term recurrence relations also give a recurrence relation of the
normalisation constants $h_n(\bm{\lambda})$.
Since \eqref{3term} and \eqref{phin_form} imply (we suppress $\bm{\lambda}$)
\begin{align*}
  &\quad\,(\phi_n,\eta\phi_{n-1})
  =(\phi_n,A_{n-1}\phi_n+B_{n-1}\phi_{n-1}+C_{n-1}\phi_{n-2})=A_{n-1}h_n\\
  &=(\eta\phi_n,\phi_{n-1})
  =(A_n\phi_{n+1}+B_n\phi_n+C_n\phi_{n-1},\phi_{n-1})=C_nh_{n-1},
\end{align*}
we obtain
\begin{equation}
  \frac{h_n(\bm{\lambda})}{h_{n-1}(\bm{\lambda})}
  =\frac{c_n(\bm{\lambda})}{c_{n-1}(\bm{\lambda})}C_n(\bm{\lambda})
  \quad(1\leq n\leq n_{\text{max}}),
  \label{hn->hn-12}
\end{equation}
namely,
\begin{equation}
  h_n(\bm{\lambda})=\frac{c_n(\bm{\lambda})}{c_0(\bm{\lambda})}
  \prod_{k=1}^nC_k(\bm{\lambda})\cdot h_0(\bm{\lambda})
  \quad(0\leq n\leq n_{\text{max}}).
  \label{hnprop2}
\end{equation}
Eqs.\,\eqref{hnprop} and \eqref{hnprop2} imply that $h_0(\bm{\lambda})$
should satisfy
\begin{equation}
  \frac{h_0(\bm{\lambda}+n\bm{\delta})}{h_0(\bm{\lambda})}
  =\frac{c_n(\bm{\lambda})}{c_0(\bm{\lambda})}
  \prod_{k=1}^n
  \frac{f_k\bigl(\bm{\lambda}+(n-k)\bm{\delta}\bigr)}
  {b_{k-1}\bigl(\bm{\lambda}+(n-k)\bm{\delta}\bigr)}
  C_k(\bm{\lambda})\quad(0\leq n\leq n_{\text{max}}).
  \label{h0prop}  
\end{equation}

The explicit forms of $h_n(\bm{\lambda})$ are conjectured as
\begin{align}
  &\text{(\romannumeral5),\,(\romannumeral6)}
  :\ \ h_n(\bm{\lambda})=2\pi\prod_{k=0}^{n-1}4\sin\tfrac{\gamma}{2}(k+1)
  \sin\gamma(n+\alpha-1-\tfrac{k}{2})\n[-3pt]
  &\phantom{\text{(\romannumeral5),\,(\romannumeral6)}:\ \ h_n(\bm{\lambda})=}
  \times\frac{\Phi_{\frac{\gamma}{2}}\bigl(i(\pi-\frac{\gamma}{2})\bigr)}
  {\Phi_{\frac{\gamma}{2}}\bigl(-i(3\pi+\gamma(2n+2\alpha-\frac12))\bigr)}
  \cdot\prod_{j=1}^2\Phi_{\frac{\gamma}{2}}
  \bigl(-i(\pi+\gamma(n+2\alpha_j-\tfrac12))\bigr)\n
  &\phantom{\text{(\romannumeral5),\,(\romannumeral6)}:\ \ h_n(\bm{\lambda})=}
  \times\prod_{\epsilon=\pm 1}
  \Phi_{\frac{\gamma}{2}}\bigl(\epsilon\gamma(\beta_1-\beta_2)
  -i(\pi+\gamma(n+\alpha-\tfrac12))\bigr)
  \label{hn(v)}\\
  &\phantom{\text{(\romannumeral5),\,(\romannumeral6)}:\ \ h_n(\bm{\lambda})=}
  \times e^{i\frac{\gamma}{2}\bigl((\alpha_1-\alpha_2)^2
  -(\beta_1-\beta_2)^2-(n+\alpha)(\frac{2\pi}{\gamma}+1)
  -\frac{8\pi^2}{3\gamma^2}-\frac{\pi}{\gamma}+\frac13\bigr)}
  \times e^{-(\beta_1+\beta_2)(\pi+\gamma(n+\alpha-\frac12))},\n
  &\text{(\romannumeral7)}
  :\ \ h_n(\bm{\lambda})=2\pi\prod_{k=0}^{n-1}4\sin\tfrac{\gamma}{2}(k+1)
  \sin\gamma(n+\alpha-1-\tfrac{k}{2})\n[-3pt]
  &\phantom{\text{(\romannumeral7)}:\ \ h_n(\bm{\lambda})=}
  \times\frac{\Phi_{\frac{\gamma}{2}}\bigl(i(\pi-\frac{\gamma}{2})\bigr)}
  {\Phi_{\frac{\gamma}{2}}\bigl(-i(3\pi+\gamma(2n+2\alpha-\frac12))\bigr)}
  \cdot\prod_{j=1}^2\Phi_{\frac{\gamma}{2}}
  \bigl(-i(\pi+\gamma(n+2\alpha_j-\tfrac12))\bigr)\n
  &\phantom{\text{(\romannumeral7)}:\ \ h_n(\bm{\lambda})=}
  \times\prod_{\epsilon_1,\epsilon_2=\pm 1}
  \Phi_{\frac{\gamma}{2}}\bigl(\gamma(\epsilon_1\beta_1+\epsilon_2\beta_2)
  -i(\pi+\gamma(n+\alpha-\tfrac12))\bigr)
  \label{hn(vii)}\\
  &\phantom{\text{(\romannumeral7)}:\ \ h_n(\bm{\lambda})=}
  \times e^{i\frac{\gamma}{2}\bigl((n+\alpha-1)^2+(\alpha_1-\alpha_2)^2
  -2(\beta_1^2+\beta_2^2)-2(\frac12+\frac{\pi}{\gamma})^2\bigr)},\n
  &\text{(\romannumeral8)}
  :\ \ h_n(\bm{\lambda})
  =2\pi\prod_{k=0}^{n-1}(-1)^{K+1}4\sin\tfrac{\gamma}{2}(k+1)
  \sin\gamma(n+\alpha-1-\tfrac{k}{2})\n[-3pt]
  &\phantom{\text{(\romannumeral8)}:\ \ h_n(\bm{\lambda})=}
  \times\frac{\Phi_{\frac{\gamma}{2}}\bigl(i(\pi-\frac{\gamma}{2})\bigr)}
  {\Phi_{\frac{\gamma}{2}}\bigl(-i((1+2K)\pi+\gamma(2n+2\alpha-\frac12))\bigr)}
  \n
  &\phantom{\text{(\romannumeral8)}:\ \ h_n(\bm{\lambda})=}
  \times\prod_{j=1}^2\Phi_{\frac{\gamma}{2}}\bigl(
  -i(-\tfrac{\alpha_j}{|\alpha_j|}\pi+\gamma(n+2\alpha_j-\tfrac12))\bigr)
  \label{hn(viii)}\\
  &\phantom{\text{(\romannumeral8)}:\ \ h_n(\bm{\lambda})=}
  \times\prod_{\epsilon=\pm 1}
  \Phi_{\frac{\gamma}{2}}\bigl(\epsilon\gamma(\beta_1-\beta_2)
  -i((1+K-K^2)\pi+\gamma(n+\alpha-\tfrac12))\bigr)\n
  &\phantom{\text{(\romannumeral8)}:\ \ h_n(\bm{\lambda})=}
  \times\prod_{\epsilon=\pm 1}
  \Phi_{\frac{\gamma}{2}}\bigl(\epsilon\gamma(\beta_1+\beta_2)
  -i(K(K+1)\pi+\gamma(n+\alpha-\tfrac12))\bigr)\n
  &\phantom{\text{(\romannumeral8)}:\ \ h_n(\bm{\lambda})=}
  \times e^{i\frac{\gamma}{2}\bigl((n+\alpha)^2
  +2(n+\alpha)(\frac{K\pi}{\gamma}-1)
  +(\alpha_1-\alpha_2)(\alpha_1-\alpha_2
  -\frac{\pi}{\gamma}(\frac{\alpha_1}{|\alpha_1|}-\frac{\alpha_2}{|\alpha_2|}))
  -2(\beta_1^2+\beta_2^2)+\frac12-\frac{\pi}{\gamma}(1+2K)
  +\frac{\pi^2}{\gamma^2}\bigr)}\n
  &\phantom{\text{(\romannumeral8)}:\ \ h_n(\bm{\lambda})=}
  \times e^{\pi(\frac{\alpha_1}{|\alpha_1|}\beta_1
  +\frac{\alpha_2}{|\alpha_2|}\beta_2)},\nonumber
\end{align}
which are supported by numerical calculation.
We can check that they satisfy the properties \eqref{hn->hn-1}--\eqref{hnprop}
and \eqref{hn->hn-12}--\eqref{h0prop}.

\subsection{Limit to oQM}
\label{sec:limit}

In an appropriate $\gamma\to 0$ limit, idQM reduces to oQM \cite{os6}.
Let us take the parameters as
\begin{align}
  \text{(\romannumeral5),\,(\romannumeral6)}:&
  \ \ \alpha_1=-\tfrac{\pi}{2\gamma}-h_1,
  \ \ \alpha_2=-\tfrac{\pi}{2\gamma}-h+h_1+\tfrac12,
  \ \ e^{\gamma\beta_j}=(\gamma\beta'_j)^{-1},\n
  \text{(\romannumeral7)}:&
  \ \ \alpha_1=-\tfrac{\pi}{\gamma}+\tfrac12(g+\tfrac12),
  \ \ \alpha_2=\tfrac12(-h+\tfrac12),\\
  \text{(\romannumeral8)}:&
  \ \ \alpha_1=-\tfrac{\pi}{2\gamma}-h_1,
  \ \ \alpha_2=-\tfrac{\pi}{2\gamma}-h+h_1+\tfrac12,\ \ K=1.
  \nonumber
\end{align}
Here we assume that $g$, $h$, $h_1$ and $\beta'_j$ in
(\romannumeral5)--(\romannumeral6) and $\beta_j$ in
(\romannumeral7)--(\romannumeral8) are independent of $\gamma$.
Note that $h_1$ is a redundant parameter and $\beta'_j>0$.
By taking $\gamma\to0$ limit of $\mathcal{A}(\bm{\lambda})$, we obtain
\begin{alignat}{2}
  \text{(\romannumeral5)}:&
  \ \ \lim_{\gamma\to0}\tfrac{1}{\gamma^2}\mathcal{H}(\bm{\lambda})
  =\mathcal{H}^{\text{M}},&&\quad\beta'_1+\beta'_2=\mu,\n
  \text{(\romannumeral6)}:&
  \ \ \lim_{\gamma\to0}\tfrac{1}{\gamma^2}\mathcal{H}(\bm{\lambda})
  \bigl|_{x\to-x}
  =\mathcal{H}^{\text{M}},&&\quad\beta'_1+\beta'_2=\mu,\n
  \text{(\romannumeral7)}:&
  \ \ \lim_{\gamma\to0}\tfrac{4}{\gamma^2}\mathcal{H}(\bm{\lambda})
  \bigl|_{x\to2x}
  =\mathcal{H}^{\text{hDPT}},
  \label{limH}\\
  \text{(\romannumeral8)}:&
  \ \ \lim_{\gamma\to0}\tfrac{1}{\gamma^2}\mathcal{H}(\bm{\lambda})
  =\mathcal{H}^{\text{hst}},&&\quad\beta_1+\beta_2=\mu.
  \nonumber
\end{alignat}
Here $\mathcal{H}^{\text{M}}$, $\mathcal{H}^{\text{hDPT}}$ and
$\mathcal{H}^{\text{hst}}$ are the Hamiltonians for Morse, hyperbolic
Darboux-P\"{o}schl-Teller and hyperbolic symmetric top $\II$
(see for example \cite{os28,os29}),
\begin{alignat}{2}
  \mathcal{H}^{\text{M}}
  &=p^2+\mu^2e^{2x}-\mu(2h+1)e^x+h^2,&-\infty<&\ x<\infty,\n
  \mathcal{H}^{\text{hDPT}}
  &=p^2+\frac{g(g-1)}{\sinh^2x}-\frac{h(h+1)}{\cosh^2 x}+(h-g)^2,
  &0<&\ x<\infty,\\
  \mathcal{H}^{\text{hst}}
  &=p^2+\frac{-h(h+1)+\mu^2+\mu(2h+1)\sinh x}{\cosh^2x}+h^2,
  &\quad-\infty<&\ x<\infty,
  \nonumber
\end{alignat}
and $\mathcal{H}^{\text{hst}}$ with $\mu=0$ is the Hamiltonian for the soliton
potential.
It is easy to show that the eigenvalues $\mathcal{E}_n(\bm{\lambda})$ have
the corresponding limits, too.
Considering the above limiting forms of the Hamiltonians \eqref{limH},
we know that the limit of the eigenfunctions $\phi_n(x;\bm{\lambda})$ should
be $\lim\limits_{\gamma\to0}\gamma^{\text{(some power)}}\times
\phi_n(x;\bm{\lambda})\propto\phi_n^{\text{oQM}}(x)$.
In fact this can be verified by direct calculation with the use of the formulas
in Appendix \ref{app:qdilog} and \ref{app:Pndet},
\begin{align}
  \text{(\romannumeral5)}:&
  \ \ \lim_{\gamma\to0}\phi_0(x;\bm{\lambda})
  =\phi_0^{\text{M}}(x),\quad
  \phi_0^{\text{M}}(x)=e^{hx-\mu e^x},\n
  &\ \ \lim_{\gamma\to0}\gamma^{-n}P_n\bigl(\eta(x);\bm{\lambda}\bigr)
  =(2\mu)^nn!\,P_n^{\text{M}}(e^{-x}),
  \ P_n^{\text{M}}(\eta)=(2\mu\eta^{-1})^{-n}L^{(2h-2n)}_n(2\mu\eta^{-1}),\n
  \text{(\romannumeral7)}:&
  \ \ \lim_{\gamma\to0}\phi_0(x;\bm{\lambda})\bigl|_{x\to2x}
  =2^{g-h}\phi_0^{\text{hDPT}}(x),\n
  &\ \ \quad
  \phi_0^{\text{hDPT}}(x)=\bigl(\tfrac{\cosh 2x-1}{2}\bigr)^{\frac{g}{2}}
  \bigl(\tfrac{\cosh 2x+1}{2}\bigr)^{-\frac{h}{2}}
  =\bigl(\sinh x\bigr)^g\bigl(\cosh x\bigr)^{-h},\n
  &\ \ \lim_{\gamma\to0}\gamma^{-n}P_n\bigl(\eta(x);\bm{\lambda}\bigr)
  \bigl|_{x\to2x}
  =(-1)^n2^{2n}n!P_n^{\text{hDPT}}(\cosh2x),\n
  &\ \ \quad
  P_n^{\text{hDPT}}(\eta)=P^{(g-\frac12,-h-\frac12)}_n(\eta),\\
  \text{(\romannumeral8)}:&
  \ \ \lim_{\gamma\to0}\phi_0(x;\bm{\lambda})
  =2^{-h}e^{-\frac12\pi\mu}\phi_0^{\text{hst}}(x),\quad
  \phi_0^{\text{hst}}(x)=\bigl(\cosh x\bigr)^{-h}e^{-\mu\tan^{-1}\sinh x},\n
  &\ \ \lim_{\gamma\to0}\gamma^{-n}P_n\bigl(\eta(x);\bm{\lambda}\bigr)
  =2^{2n}n!\,P_n^{\text{hst}}(\sinh x),
  \ P_n^{\text{hst}}(\eta)
  =i^{-n}P^{(-h-\frac12-i\mu,-h-\frac12+i\mu)}_n(i\eta).
  \nonumber
\end{align}
Here $L^{(\alpha)}_n(\eta)$ and $P_n^{(\alpha,\beta)}(\eta)$ are the Laguerre
and the Jacobi polynomials in $\eta$ of degree $n$, respectively.
The limit of $n_{\text{max}}(\bm{\lambda})$ also gives the corresponding one.
The case (\romannumeral6) can be obtained from (\romannumeral5) by the
replacement $x\to-x$.

\subsection{Other limits}
\label{sec:limit2}

It is well known that the idQM system of the Wilson polynomial can be
obtained from that of the Askey-Wilson polynomial by the replacements
$x^{\text{AW}}=\frac{\pi}{L}x^{\text{W}}$ and
$\gamma^{\text{AW}}=-\frac{\pi}{L}$ and taking $L\to\infty$
($q=e^{\gamma^{\text{AW}}}\to 1$) limit \cite{os13}.
Let us consider similar $q\to 1$ limits of (\romannumeral7) and
(\romannumeral6) cases.

Let us set
\begin{equation}
  x=\frac{x'}{R},\quad\gamma=\frac{1}{R},\quad
  \lambda_j=-\pi R+\lambda'_j\ \ (j=1,2),
\end{equation}
where $R>0$ and we assume that $\lambda'_j$ is $R$-independent.
In the $R\to\infty$ ($q=e^{-i\gamma}\to 1$) limit we have
\begin{align}
  \text{(\romannumeral7)}:&\quad
  \lim_{R\to\infty}R^2\,V(x;\bm{\lambda})
  =\frac{\prod_{j=1}^2(\lambda'_j+ix')(\lambda^{\prime\,*}_j+ix')}
  {2ix'(2ix'+1)}
  =V^{\text{W}}(x';\bm{\lambda}^{\text{W}}),\n
  \text{(\romannumeral6)}:&\quad
  \lim_{R\to\infty}R^2\,V(x;\bm{\lambda})
  =\prod_{j=1}^2(\lambda'_j+ix')
  =V^{\text{cH}}(x';\bm{\lambda}^{\text{cH}}).
\end{align}
Here $V^{\text{W}}$ and $V^{\text{cH}}$ are the potential functions of the
Wilson and the continuous Hahn systems, respectively and
$\bm{\lambda}^{\text{W}}=(\lambda'_1,\lambda'_2,\lambda^{\prime\,*}_1,
\lambda^{\prime\,*}_2)$ and $\bm{\lambda}^{\text{cH}}=(\lambda'_1,\lambda'_2)$
\cite{os13}.
Thus the Wilson and the continuous Hahn systems are obtained from
(\romannumeral7) and (\romannumeral6), respectively.
We can verify the limits of the corresponding eigenpolynomials
\begin{align}
  \text{(\romannumeral7)}:&\quad
  \lim_{R\to\infty}R^{3n}
  P_n\bigl(\eta(x);\bm{\lambda}\bigr)
  =(-1)^nW_n(x^{\prime\,2};\lambda'_1,\lambda'_2,\lambda^{\prime\,*}_1,
  \lambda^{\prime\,*}_2),\n
  \text{(\romannumeral6)}:&\quad
  \lim_{R\to\infty}R^{2n}
  P_n\bigl(\eta(x);\bm{\lambda}\bigr)
  =n!\,p_n(x';\lambda'_1,\lambda'_2,\lambda^{\prime\,*}_1,
  \lambda^{\prime\,*}_2),\\
  \text{(\romannumeral7)\,(\romannumeral6)}:&\quad
  \lim_{R\to\infty}R^2
  \mathcal{E}_n(\bm{\lambda})
  =n(n+\lambda'_1+\lambda'_2+\lambda^{\prime\,*}_1+\lambda^{\prime\,*}_2-1),
  \quad
  \lim_{R\to\infty}n_{\text{max}}(\bm{\lambda})=\infty,\nonumber
\end{align}
where $W_n(\eta;a_1,a_2,a_3,a_4)$ and $p_n(\eta;a_1,a_2,a_3,a_4)$ are
the Wilson and the continuous Hahn polynomials, respectively \cite{koeswart}.

The $q$-deformation of the continuous Hahn polynomial is known as the
continuous $q$-Hahn polynomial $p_n(\eta;a_1,a_2,a_3,a_4;q)$
($-1<\eta<1$, $0<q<1$) which is the Askey-Wilson polynomial with different
parameters \cite{koeswart}.
Our $\tilde{p}_n(\eta;a_1,a_2,a_3,a_4|q)$ ($0<\eta<\infty$, $|q|=1$)
\eqref{ptn} gives another $q$-deformation of the continuous Hahn polynomial.

\section{Summary and Comments}
\label{summary}

Several kinds of $q$-orthogonal polynomials with $|q|=1$ have been constructed.
Their weight functions consist of products of quantum dilogarithm functions,
which are natural generalisation of the Euler gamma functions and $q$-gamma
functions. The total number of mutually orthogonal polynomials is finite.

In other words, several new examples of exactly solvable systems in discrete
quantum mechanics with pure imaginary shifts have been derived based on the
sinusoidal coordinates $\eta(x)=\cosh x,e^{\pm x},\sinh x$.
The method called ``unified theory of exactly and quasi-exactly solvable
discrete quantum mechanics'' was developed by the present authors several
years ago \cite{os14}.
These new systems have finitely many discrete eigenstates.
Their eigenpolynomials are given by the Askey-Wilson polynomial and its
certain limiting forms with the parameter $q=e^{-i\gamma}$, $|q|=1$.
The groundstate wavefunctions are described by the quantum dilogarithm.
In appropriate limits they reduce to known solvable systems in the ordinary
quantum mechanics: Morse, hyperbolic Darboux-P\"{o}schl-Teller and hyperbolic
symmetric top $\II$ potentials, which also have finitely many discrete
eigenstates.

Several comments are in order.
The listed parameter ranges of these new systems, \eqref{range(vii)},
\eqref{range(v)} and \eqref{range(viii)}, are rather conservative.
These could be extended with scrutiny.
We have considered the factorisations \eqref{Vtfactor}, but
$\widetilde{V}(x)$ \eqref{exV} may have different factorised forms for less
generic parameters.
For example, the following factorisation is allowed for (\romannumeral7):
\begin{equation}
  \widetilde{V}(x)=e^{-2x}A\prod_{j=1}^{4}(e^x+e^{i\theta_j})\quad
  (A>0,\ \theta_j\in\mathbb{R}).
  \label{Vt(vii)2}
\end{equation}
The four parameters $(a_1,a_2,a_3,a_4)$ of the Askey-Wilson polynomial
$p_n(\cos x;a_1,a_2,a_3,a_4|q)$ with $0<q<1$ and $|a_j|<1$ are either
(a) $a_1,a_2\in\mathbb{C}$, $a_3=a_1^*$, $a_4=a_2^*$
or (b) $a_1,a_2,a_3,a_4\in\mathbb{R}$ \cite{koeswart}.
Our factorisation \eqref{Vtfactor} corresponds to (a) and \eqref{Vt(vii)2}
corresponds to (b).
It is an interesting problem to clarify whether the well-defined quantum
systems are obtained for such less generic cases as \eqref{Vt(vii)2}.
By construction, the $q$-value of the present theory is limited as
$q=e^{-i\gamma}$, $0<\gamma<\pi$.
Recently there are interesting developments related to $q$-orthogonal
polynomials with $q=-1$ \cite{q=-1}.
So far, we have not been able to extend our theory to $q=-1$ or $\gamma=\pi$,
for which the integral representation of the quantum dilogarithm
$\Phi_{\gamma}(z)$ \eqref{intrep} has double poles.
The fact that the new solvable systems have only finitely many discrete
eigenstates mean that scattering problems can be formulated.
It is an interesting challenge to calculate the reflection and transmission
amplitudes of these systems.
The Hamiltonians of the known exactly solvable idQM systems, {\em e.g.}
the Wilson (\romannumeral2) and the Askey-Wilson (\romannumeral3) have
discrete symmetries, which are essential for multi-indexed deformations
\cite{os27,os17,os20} for these systems through Darboux transformations
\cite{os15,gos}.
It is interesting to find discrete symmetries and (psudo)virtual state
wavefunctions for (\romannumeral5)-(\romannumeral8).

After completing this work, we were informed by R.\,Askey of the sieved
orthogonal polynomials \cite{aaa,svpoly}. They could be considered as
$q$-orthogonal polynomials with $q$ being a root of unity, as they are
obtained from $q$-orthogonal polynomials, {\em e.g.} $q$-ultraspherical
polynomials \cite{askey,ismail,koeswart,rogers}, by certain limiting
procedures.

\section*{Acknowledgements}
S.\,O. thanks M.\,Jimbo for useful discussion of quantum dilogarithm.
We thank R.\,Askey for enlightening discussion on sieved orthogonal polynomials.
S.\,O. and R.\,S. are supported in part by Grant-in-Aid for Scientific Research
from the Ministry of Education, Culture, Sports, Science and Technology
(MEXT), No.25400395 and No.22540186, respectively.

\bigskip
\appendix
\section{Various Data}
\label{app:data}

\subsection{Matrix elements $\widetilde{\mathcal{H}}^{\eta}_{m,n}$}
\label{app:Htmn}

The sinusoidal coordinate $\eta(x)$ satisfying \eqref{eta_os14} has the
property (\U 2.21),
\begin{equation}
  \frac{\eta(x-i\gamma)^{n+1}-\eta(x+i\gamma)^{n+1}}
  {\eta(x-i\gamma)-\eta(x+i\gamma)}
  =\sum_{k=0}^ng_n^{(k)}\eta(x)^{n-k}\quad(n\in\mathbb{Z}_{\geq 0}),
  \label{gnk}
\end{equation}
where $g_n^{(k)}$ are real constants and we set $g_n^{(k)}=0$ unless
$0\leq k\leq n$.
The matrix elements $\widetilde{\mathcal{H}}^{\eta}_{m,n}$ \eqref{Htact}
are given by (\U 2.35)--(\U 2.36):
\begin{equation}
  \widetilde{\mathcal{H}}_{m,n}^{\eta}=\!\!\!\!
  \sum_{j=\max(n-2-m,0)}^{n-m}\!\!\!\!e_{n-m,j,n},\quad
  e_{m,j,n}\eqdef\sum_{l=0}^{2-m+j}\!\!v_{2-m+j-l,l}
  \sum_{r=0}^{n-1}g_{n+l-r-2}^{(j)}.
  \label{Htmn}
\end{equation}

\subsection{Explicit forms of $g_n^{(k)}$}
\label{app:gnk}

For the eight sinusoidal coordinates \eqref{etalist_os14} with
\eqref{gamma_range} and \eqref{r11}, the explicit forms of $[\![n]\!]$
(which was denoted as $[n]$ in (\U 2.25)) and $g_n^{(k)}$ in \eqref{gnk}
are the following:\\
\underline{$[\![n]\!]$} :
\begin{equation}
  [\![n]\!]=
  \begin{cases}
   n&\text{for (\romannumeral1)--(\romannumeral2)}\\
   {\displaystyle \frac{e^{-\gamma n}-e^{\gamma n}}
   {e^{-\gamma}-e^{\gamma}}}
   &\text{for (\romannumeral3)--(\romannumeral4)}\\[6pt]
   {\displaystyle \frac{e^{i\gamma n}-e^{-i\gamma n}}
   {e^{i\gamma}-e^{-i\gamma}}}
   &\text{for (\romannumeral5)--(\romannumeral8)}
  \end{cases}.
  \label{[[n]]}
\end{equation}
\underline{$g_n^{(k)}$} :
\begin{align}
  \text{(\romannumeral1)}:&\quad g_n^{(k)}=
  \theta(k:\text{even})\,(-1)^{\frac{k}{2}}\genfrac{(}{)}{0pt}{}{n+1}{k+1},\n
  \text{(\romannumeral2)}:&\quad g_n^{(k)}=
  \frac{(-1)^k}{2}\genfrac{(}{)}{0pt}{}{2n+2}{2k+1},\n
  \text{(\romannumeral5),\,(\romannumeral6)}:&\quad g_n^{(k)}=
  [\![n+1]\!]\,\delta_{k\,0},
  \label{ex_gnk}\\[-4pt]
  \text{(\romannumeral3),\,(\romannumeral4),\,(\romannumeral7)}:&
  \quad g_n^{(k)}=
  \theta(k:\text{even})\,\frac{(n+1)!}{2^k}
  \sum_{r=0}^{\frac{k}{2}}\genfrac{(}{)}{0pt}{}{n-k+r}{r}
  \frac{(-1)^r\,[\![n-k+1+2r]\!]}{(\frac{k}{2}-r)!\,(n-\frac{k}{2}+1+r)!},\n
  \text{(\romannumeral8)}:&\quad g_n^{(k)}=
  (-1)^{\frac{k}{2}}\times\bigl(\text{RHS of the above equation}\bigr).
  \nonumber
\end{align}
Here $\theta(P)$ is a step function for a proposition $P$ ; $\theta(P)=1$
for $P$\,:\,true, $\theta(P)=0$ for $P$\,:\,false.

\subsection{Explicit forms of $V(x)$}
\label{app:V}

For the eight sinusoidal coordinates \eqref{etalist_os14} with
\eqref{gamma_range}, the potential function $V(x)$
\eqref{V=Vt/etaeta}--\eqref{Vt} becomes
\begin{align}
  \text{(\romannumeral1)}:&
  \ V(x)=-\tfrac12\widetilde{V}(x),
  \  \widetilde{V}(x)=\sum_{k=0}^2\tilde{v}_kx^k,
  \ \tilde{v}_2\in\mathbb{R},
  \ \tilde{v}_k\in\mathbb{C}\ (k\neq 2),\n
  \text{(\romannumeral2)}:&
  \ V(x)=\frac{\tfrac12\widetilde{V}(x)}{2ix(2ix+1)},
  \ \widetilde{V}(x)=\sum_{k=0}^{4}\tilde{v}_kx^k,
  \ i^k\tilde{v}_k\in\mathbb{R},\n
  \text{(\romannumeral3)}:&
  \ V(x)=\frac{e^{\frac{\gamma}{2}}}{\sinh\frac{\gamma}{2}\sinh\gamma}
  \frac{e^{2ix}\,\widetilde{V}(x)}{(1-e^{2ix})(1-e^{\gamma}e^{2ix})},
  \ \widetilde{V}(x)=\sum_{k=-2}^2\tilde{v}_ke^{ikx},
  \ \tilde{v}_k\in\mathbb{R},\n
  \text{(\romannumeral4)}:&
  \ V(x)=\frac{-e^{\frac{\gamma}{2}}}{\sinh\frac{\gamma}{2}\sinh\gamma}
  \frac{e^{2ix}\,\widetilde{V}(x)}{(1+e^{2ix})(1+e^{\gamma}e^{2ix})},
  \ \widetilde{V}(x)=\sum_{k=-2}^2\tilde{v}_ke^{ikx},
  \ i^k\tilde{v}_k\in\mathbb{R},\n
  \text{(\romannumeral5)}:&
  \ V(x)=\frac{-e^{-i\frac{\gamma}{2}}}{4\sin\frac{\gamma}{2}\sin\gamma}\,
  e^{2x}\,\widetilde{V}(x),
  \ \widetilde{V}(x)=\sum_{k=0}^2\tilde{v}_ke^{-kx},
  \ \tilde{v}_0\in\mathbb{R},
  \ \tilde{v}_k\in\mathbb{C}\ (k\neq 0),
  \label{exV}\\
  \text{(\romannumeral6)}:&
  \ V(x)=\frac{-e^{i\frac{\gamma}{2}}}{4\sin\frac{\gamma}{2}\sin\gamma}\,
  e^{-2x}\,\widetilde{V}(x),
  \ \widetilde{V}(x)=\sum_{k=0}^2\tilde{v}^*_ke^{kx},
  \ \tilde{v}_0\in\mathbb{R},
  \ \tilde{v}_k\in\mathbb{C}\ (k\neq 0),\n
  \text{(\romannumeral7)}:&
  \ V(x)=\frac{-e^{-i\frac{\gamma}{2}}}{\sin\frac{\gamma}{2}\sin\gamma}\,
  \frac{e^{2x}\,\widetilde{V}(x)}{(1-e^{2x})(1-e^{-i\gamma}e^{2x})},
  \ \widetilde{V}(x)=\sum_{k=-2}^2\tilde{v}_ke^{kx},
  \ \tilde{v}_{-k}=\tilde{v}_k^*\in\mathbb{C},\n
  \text{(\romannumeral8)}:&
  \ V(x)=\frac{-e^{-i\frac{\gamma}{2}}}{\sin\frac{\gamma}{2}\sin\gamma}\,
  \frac{e^{2x}\,\widetilde{V}(x)}{(1+e^{2x})(1+e^{-i\gamma}e^{2x})},
  \ \widetilde{V}(x)=\sum_{k=-2}^2\tilde{v}_ke^{kx},
  \ \tilde{v}_{-k}=(-1)^k\tilde{v}_k^*\in\mathbb{C}.
  \nonumber
\end{align}
The transformation from $\{v_{k,l}\ (k+l\leq 2)\}$ to
$\{\text{Re}\,\tilde{v}_k,\text{Im}\,\tilde{v}_k\}$ is a real linear
map of rank $5$.
The condition $\sum\limits_{k+l=2}v_{k,l}^2\neq 0$ becomes:
(\romannumeral1): $(\tilde{v}_2,\text{Im}\,\tilde{v}_1)\neq(0,0)$,
(\romannumeral2): $(\tilde{v}_4,\tilde{v}_3)\neq(0,0)$,
(\romannumeral3)--(\romannumeral4): $(\tilde{v}_2,\tilde{v}_{-2})\neq(0,0)$,
(\romannumeral5)--(\romannumeral8): $\tilde{v}_2\neq 0$.

\subsection{Factorised forms of $\widetilde{V}(x)$}
\label{app:facV}

For the generic values of the original parameters $\{v_{k,l}\}$, the above
$\widetilde{V}(x)$ have the following factorised forms:
\begin{align}
  \text{(\romannumeral1)}:&
  \ \widetilde{V}(x)=(-i)^2A\prod_{j=1}^2(a_j+ix),\quad
  \ \ \text{(\romannumeral2)}:
  \ \widetilde{V}(x)=A\prod_{j=1}^{4}(a_j+ix),\n
  \text{(\romannumeral3)}:&
  \ \widetilde{V}(x)=e^{-i2x}A\prod_{j=1}^{4}(1-a_je^{ix}),\quad
  \text{(\romannumeral4)}:
  \ \widetilde{V}(x)=(-i)^2e^{-i2x}A\prod_{j=1}^{4}(1-ia_je^{ix}),\n
  \text{(\romannumeral5)}:&
  \ \widetilde{V}(x)=A\prod_{j=1}^2(1+a_j^*e^{-x}),\quad
  \quad\ \ \,\text{(\romannumeral6)}:
  \ \widetilde{V}(x)=A\prod_{j=1}^2(1+a_je^x),
  \label{Vtfactor}\\
  \text{(\romannumeral7)}:&
  \ \widetilde{V}(x)=A\prod_{j=1}^2(1+a_je^x)(1+a_j^*e^{-x}),\quad
  \text{(\romannumeral8)}:
  \ \widetilde{V}(x)=A\prod_{j=1}^2(1+a_je^x)(1-a_j^*e^{-x}),
  \nonumber
\end{align}
where $A$ is a real overall scaling parameter and $a_j\in\mathbb{C}$ are
the new parameters.
For (\romannumeral2)--(\romannumeral4), the four parameters $a_j$ obey
$\{a_1^*,\ldots,a_{4}^*\}=\{a_1,\ldots,a_{4}\}$ (as a set).
For less generic cases, $\widetilde{V}(x)$ may have different factorised forms.

We present the relations among the original real parameters $v_{k,l}$ in
$V(x)$ \eqref{V=Vt/etaeta}--\eqref{Vt} and the new complex parameters
$a_j$ after factorisation in \eqref{Vtfactor}.
Since $v_{0,2}$ is redundant (see \eqref{eta(x-ig)^2}), we set $v_{0,2}=0$.
Then $v_{j,k}$ are expressed in terms of $a_j$ and the overall scaling
parameter $A$:
\begin{align}
  \text{(\romannumeral1)}:&
  \ v_{0,0}=-A\,\text{Re}(a_1a_2),
  \ v_{0,1}=A\,\text{Im}(a_1a_2),
  \ v_{1,0}=A\,\text{Im}(a_1+a_2-a_1a_2),\n
  &\ v_{1,1}=A\,\text{Re}(a_1+a_2),
  \ v_{2,0}=A\bigl(1-\text{Re}(a_1+a_2)\bigr),\n
  \text{(\romannumeral2)}:&
  \ v_{0,0}=-\tfrac12A(b_3-2b_4),
  \ v_{0,1}=-\tfrac12Ab_3,
  \ v_{1,0}=\tfrac12A(b_1-2b_2+b_3),
  \ v_{1,1}=\tfrac12Ab_1,\n
  &\ v_{2,0}=\tfrac12A(2-b_1),\n
  \text{(\romannumeral3)}:&
  \ v_{0,0}=-A(1-b_2+b_4),
  \ v_{0,1}=\frac{-2A(b_1-b_3)}{e^{-\gamma}-e^{\gamma}},
  \ v_{1,0}=\frac{2A(e^{\gamma}b_1-e^{-\gamma}b_3)}{e^{-\gamma}-e^{\gamma}},\n
  &\ v_{1,1}=\frac{4A(1-b_4)}{e^{-\gamma}-e^{\gamma}},
  \ v_{2,0}=\frac{-4A(e^{\gamma}-e^{-\gamma}b_4)}{e^{-\gamma}-e^{\gamma}},\n
  \text{(\romannumeral4)}:&
  \ v_{j,k}=(-1)^{j+k}v_{j,k}^{\text{(\romannumeral3)}},\n
  \text{(\romannumeral5),\,(\romannumeral6)}:&
  \ v_{0,0}=A,
  \ v_{0,1}=\frac{-A(a_1+a_2-a_1^*-a_2^*)}{e^{i\gamma}-e^{-i\gamma}},
  \ v_{1,0}=\frac{A\bigl(e^{i\gamma}(a_1+a_2)-e^{-i\gamma}(a_1^*+a_2^*)\bigr)}
  {e^{i\gamma}-e^{-i\gamma}},\n
  &\ v_{1,1}=\frac{-A(a_1a_2-a_1^*a_2^*)}{e^{i\gamma}-e^{-i\gamma}},
  \ v_{2,0}=\frac{A(e^{i\gamma}a_1a_2-e^{-i\gamma}a_1^*a_2^*)}
  {e^{i\gamma}-e^{-i\gamma}},
  \label{vjk-aj}\\
  \text{(\romannumeral7)}:&
  \ v_{0,0}=A\bigl((1-a_1a_2)(1-a_1^*a_2^*)+(a_1+a_2)(a_1^*+a_2^*)\bigr),\n
  &\ v_{0,1}=\frac{-2A\bigl((a_1-a_1^*)(1+a_2a_2^*)
  +(a_2-a_2^*)(1+a_1a_1^*)\bigr)}{e^{i\gamma}-e^{-i\gamma}},\n
  &\ v_{1,0}=\frac{2A\bigl(e^{i\gamma}(a_1(1+a_2a_2^*)+a_2(1+a_1a_1^*))
  -e^{-i\gamma}(a_1^*(1+a_2a_2^*)+a_2^*(1+a_1a_1^*))\bigr)}
  {e^{i\gamma}-e^{-i\gamma}},\n
  &\ v_{1,1}=\frac{-4A(a_1a_2-a_1^*a_2^*)}{e^{i\gamma}-e^{-i\gamma}},
  \ v_{2,0}=\frac{4A(e^{i\gamma}a_1a_2-e^{-i\gamma}a_1^*a_2^*)}
  {e^{i\gamma}-e^{-i\gamma}},\n
  \text{(\romannumeral8)}:&
  \ v_{0,0}=A\bigl((1+a_1a_2)(1+a_1^*a_2^*)-(a_1+a_2)(a_1^*+a_2^*)\bigr),\n
  &\ v_{0,1}=\frac{-2A\bigl((a_1-a_1^*)(1-a_2a_2^*)
  +(a_2-a_2^*)(1-a_1a_1^*)\bigr)}{e^{i\gamma}-e^{-i\gamma}},\n
  &\ v_{1,0}=\frac{2A\bigl(e^{i\gamma}(a_1(1-a_2a_2^*)+a_2(1-a_1a_1^*))
  -e^{-i\gamma}(a_1^*(1-a_2a_2^*)+a_2^*(1-a_1a_1^*))\bigr)}
  {e^{i\gamma}-e^{-i\gamma}},\n
  &\ v_{1,1}=\frac{-4A(a_1a_2-a_1^*a_2^*)}{e^{i\gamma}-e^{-i\gamma}},
  \ v_{2,0}=\frac{4A(e^{i\gamma}a_1a_2-e^{-i\gamma}a_1^*a_2^*)}
  {e^{i\gamma}-e^{-i\gamma}},
  \nonumber
\end{align}
where $b_j$'s in (\romannumeral2)--(\romannumeral4) are
\begin{equation*}
  b_1\eqdef\sum_{j=1}^4a_j,
  \ b_2\eqdef\sum_{1\leq j<k\leq 4}a_ja_k,
  \ b_3\eqdef\sum_{1\leq j<k<l\leq 4}a_ja_ka_l,
  \ b_4\eqdef a_1a_2a_3a_4.
\end{equation*}

\subsection{Explicit forms of determinant \eqref{Pndet} (eigenpolynomial)}
\label{app:Pndet}

For the eight sinusoidal coordinates \eqref{etalist_os14} with
\eqref{gamma_range}, the determinant expression of the eigenpolynomials
\eqref{Pndet} can be evaluated explicitly in terms of the factorisation
parameters $\{a_j\}$:
\begin{align}
  &\Bigl(\frac{A}{2}\Bigr)^n(n!)^2\times
  p_n\bigl(\eta(x)\,;a_1,a_2,a_1^*,a_2^*\bigr)
  \quad:\text{(\romannumeral1)},\n
  &\Bigl(\frac{-A}{2}\Bigr)^nn!\times
  W_n\bigl(\eta(x)\,;a_1,a_2,a_3,a_4\bigr)
  \quad:\text{(\romannumeral2)},\n
  &\biggl(\frac{A}{\sinh\frac{\gamma}{2}\sinh\gamma}\biggr)^n\,
  \prod_{k=1}^n\sinh\tfrac{-k\gamma}{2}\cdot e^{-\frac34\gamma n(n-1)}\,
  \times\left\{
  \begin{array}{ll}
  p_n\bigl(\eta(x)\,;a_1,a_2,a_3,a_4|e^{\gamma}\bigr)
  &:\text{(\romannumeral3)}\\[2pt]
  (-1)^np_n\bigl(-\eta(x)\,;a_1,a_2,a_3,a_4|e^{\gamma}\bigr)
  &:\text{(\romannumeral4)}
  \end{array}\right.\!\!,\n
  &\biggl(\frac{A|a_1a_2|}{\sin\frac{\gamma}{2}\sin\gamma}\biggr)^n\,
  \prod_{k=1}^n\sin\tfrac{k\gamma}{2}
  \times e^{-i\frac{\pi}{2}n}e^{i\frac34\gamma n(n-1)}
  \Bigl(\frac{a_1^*a_2^*}{|a_1a_2|}\Bigr)^n
  \label{exPndet}\\
  &\qquad\qquad\times\left\{
  \begin{array}{ll}
  \tilde{p}_n\bigl(\eta(x)\,;
  -a_1,-a_2,-a_1^{*\,-1},-a_2^{*\,-1}|e^{-i\gamma}\bigr)\times 2^{-n}
  &:\text{(\romannumeral5),\,(\romannumeral6)}\\[2pt]
  p_n\bigl(\eta(x)\,;
  -a_1,-a_2,-a_1^{*\,-1},-a_2^{*\,-1}|e^{-i\gamma}\bigr)
  &:\text{(\romannumeral7)}\\[2pt]
  (-i)^np_n\bigl(i\eta(x)\,;
  ia_1,ia_2,-ia_1^{*\,-1},-ia_2^{*\,-1}|e^{-i\gamma}\bigr)
  &:\text{(\romannumeral8)}
  \end{array}\right.\!\!\!.\nonumber
\end{align}
Here $p_n(\eta;a_1,a_2,a_3,a_4)$ in (\romannumeral1) and
$W_n(\eta;a_1,a_2,a_3,a_4)$ in (\romannumeral2) are the continuous Hahn and
the Wilson polynomials respectively, and
$p_n(\eta;a_1,a_2,a_3,a_4|q)$ in (\romannumeral3)--(\romannumeral4) and
(\romannumeral7)--(\romannumeral8) is the Askey-Wilson polynomial
\cite{koeswart},
\begin{align}
  &\quad p_n(\cos x;a_1,a_2,a_3,a_4|q)\n
  &\eqdef a_1^{-n}(a_1a_2,a_1a_3,a_1a_4;q)_n
  \ {}_4\phi_3\Bigl(\genfrac{}{}{0pt}{}
  {q^{-n},\,a_1a_2a_3a_4q^{n-1},\,a_1e^{ix},\,a_1e^{-ix}}
  {a_1a_2,\,a_1a_3,\,a_1a_4}\!\!\Bigm|\!q\,;q\Bigr),
  \label{AWpoly} 
\end{align}
where the basic hypergeometric series ${}_r\phi_s$ is
\begin{align}
   &{}_r\phi_s\Bigl(
   \genfrac{}{}{0pt}{}{a_1,\,\cdots,a_r}{b_1,\,\cdots,b_s}
   \Bigm|q\,;z\Bigr)
   \eqdef\sum_{n=0}^{\infty}
   \frac{(a_1,\,\cdots,a_r\,;q)_n}{(b_1,\,\cdots,b_s\,;q)_n}
   (-1)^{(1+s-r)n}q^{\frac12(1+s-r)n(n-1)}\frac{z^n}{(q\,;q)_n}\,,\n
   &\qquad(a_1,\,\cdots,a_r\,;q)_n\eqdef\prod_{j=1}^r(a_j\,;q)_n,\quad
   (a\,;q)_n\eqdef\prod_{k=1}^n(1-aq^{k-1}).
   \label{defqhypergeom}
\end{align}
The polynomial $\tilde{p}_n(\eta;a_1,a_2,a_3,a_4|q)$ in
(\romannumeral5)--(\romannumeral6) is defined as a limit of the Askey-Wilson
polynomial:
\begin{align}
  \tilde{p}_n(\eta;a_1,a_2,a_3,a_4|q)&\eqdef
  \lim_{t\to0}\,t^np_n\bigl(\tfrac{\eta}{2t};ta_1,ta_2,\tfrac{a_3}{t},
  \tfrac{a_4}{t}|q)\n
  &=a_1^{-n}(a_1a_3,a_1a_4;q)_n
  \ {}_3\phi_2\Bigl(\genfrac{}{}{0pt}{}{q^{-n},\,a_1a_2a_3a_4q^{n-1},\,a_1\eta}
  {a_1a_3,\,a_1a_4}\!\!\Bigm|\!q\,;q\Bigr),
  \label{ptn}
\end{align}
which can be regarded as a $q$-deformation of the continuous Hahn polynomial
$p_n(\eta;a_1,a_2,a_3,$ $a_4)$ \cite{koeswart},
\begin{equation*}
  p_n(\eta;a_1,a_2,a_3,a_4)
  =i^n\frac{(a_1+a_3,a_1+a_4)_n}{n!}
  \ {}_3F_2\Bigl(\genfrac{}{}{0pt}{}{-n,\,n+a_1+a_2+a_3+a_4-1,\,a_1+i\eta}
  {a_1+a_3,\,a_1+a_4}\!\!\Bigm|\!1\Bigr).
\end{equation*}
The corresponding energy eigenvalues $\mathcal{E}_n$ are
\begin{align}
  \text{(\romannumeral1)}&:\quad
  \mathcal{E}_n=\tfrac12A\times n(n+a_1+a_2+a_1^*+a_2^*-1),\n
  \text{(\romannumeral2)}&:\quad
  \mathcal{E}_n=\tfrac12A\times n(n+a_1+a_2+a_3+a_4-1),\n
  \text{(\romannumeral3),\,(\romannumeral4)}&:\quad
  \mathcal{E}_n=\frac{Ae^{\frac12\gamma}}{\sinh\frac{\gamma}{2}\sinh\gamma}
  \times(e^{-\gamma n}-1)(1-a_1a_2a_3a_4e^{\gamma(n-1)}),
  \label{exEn}\\
  \text{(\romannumeral5),\,(\romannumeral6)}&:\quad
  \mathcal{E}_n=\frac{A|a_1a_2|}{4\sin\frac{\gamma}{2}\sin\gamma}
  \times 4\sin\tfrac{\gamma}{2}n\cdot
  \frac{1}{2i}\Bigl(e^{i\frac{\gamma}{2}(n-1)}\frac{|a_1a_2|}{a_1a_2}
  -e^{-i\frac{\gamma}{2}(n-1)}\frac{|a_1a_2|}{a_1^*a_2^*}\Bigr),\n
  \text{(\romannumeral7),\,(\romannumeral8)}&:\quad
  \mathcal{E}_n=\frac{A|a_1a_2|}{\sin\frac{\gamma}{2}\sin\gamma}
  \times 4\sin\tfrac{\gamma}{2}n\cdot
  \frac{1}{2i}\Bigl(e^{i\frac{\gamma}{2}(n-1)}\frac{|a_1a_2|}{a_1a_2}
  -e^{-i\frac{\gamma}{2}(n-1)}\frac{|a_1a_2|}{a_1^*a_2^*}\Bigr).
  \nonumber
\end{align}

\medskip

We present some properties of the Askey-Wilson polynomial $p_n$ \cite{koeswart}
and $\tilde{p}_n$ :
\begin{align}
  &p_n(\eta;a_1,a_2,a_3,a_4|q):\text{ symmetric in $(a_1,a_2,a_3,a_4)$},
  \label{pnprop1}\\
  &p_n(\eta;a_1,a_2,a_3,a_4|q^{-1})=(-1)^n(a_1a_2a_3a_4)^nq^{-\frac32n(n-1)}
  p_n(\eta;a_1^{-1},a_2^{-1},a_3^{-1},a_4^{-1}|q),
  \label{pnprop2}\\
  &p_n(-\eta;a_1,a_2,a_3,a_4|q)=(-1)^np_n(\eta;-a_1,-a_2,-a_3,-a_4|q),
  \label{pnprop3}\\
  &\tilde{p}_n(\eta;a_1,a_2,a_3,a_4|q):\text{ symmetric under
  $a_1\leftrightarrow a_2$ or $a_3\leftrightarrow a_4$},
  \label{ptnprop1}\\
  &\tilde{p}_n(\eta;a_1,a_2,a_3,a_4|q^{-1})
  =(-1)^n(a_1a_2a_3a_4)^nq^{-\frac32n(n-1)}
  \tilde{p}_n(\eta;a_3^{-1},a_4^{-1},a_1^{-1},a_2^{-1}|q),
  \label{ptnprop2}\\
  &p_n(\eta;a_1,a_2,a_3,a_4|q)=c_n\eta^n+(\text{lower order terms}),\quad
  c_n=2^n(a_1a_2a_3a_4q^{n-1};q)_n,
  \label{pncn}\\
  &\tilde{p}_n(\eta;a_1,a_2,a_3,a_4|q)
  =\tilde{c}_n\eta^n+(\text{lower order terms}),\quad
  \tilde{c}_n=(a_1a_2a_3a_4q^{n-1};q)_n.
  \label{ptncn}
\end{align}
The three term recurrence relations of $p_n$ and $\tilde{p}_n$ are
\begin{align}
  &\eta p_n(\eta)=A_np_{n+1}(\eta)+B_np_n(\eta)+C_np_{n-1}(\eta),\quad
  p_n(\eta)=p_n(\eta;a_1,a_2,a_3,a_4|q),\n
  &\quad A_n=\frac{1-b_4q^{n-1}}{2(1-b_4q^{2n-1})(1-b_4q^{2n})},\quad
  b_4=a_1a_2a_3a_4,
  \label{pn3term}\\
  &\quad B_n=\frac{a_1+a_1^{-1}}{2}
  -\frac{(1-b_4q^{n-1})\prod_{j=2}^4(1-a_1a_jq^n)}
  {2a_1(1-b_4q^{2n-1})(1-b_4q^{2n})}
  -\frac{a_1(1-q^n)\prod_{2\leq j<k\leq 4}(1-a_ja_kq^{n-1})}
  {2(1-b_4q^{2n-2})(1-b_4q^{2n-1})},\n
  &\quad C_n=\frac{(1-q^n)\prod_{1\leq j<k\leq 4}(1-a_ja_kq^{n-1})}
  {2(1-b_4q^{2n-2})(1-b_4q^{2n-1})},\n
  &\eta\tilde{p}_n(\eta)=\tilde{A}_n\tilde{p}_{n+1}(\eta)
  +\tilde{B}_n\tilde{p}_n(\eta)+\tilde{C}_n\tilde{p}_{n-1}(\eta),\quad
  \tilde{p}_n(\eta)=\tilde{p}_n(\eta;a_1,a_2,a_3,a_4|q),\n
  &\quad\tilde{A}_n=\frac{1-b_4q^{n-1}}{(1-b_4q^{2n-1})(1-b_4q^{2n})},\quad
  b_4=a_1a_2a_3a_4,
  \label{ptn3term}\\
  &\quad\tilde{B}_n=a_1^{-1}-\frac{(1-b_4q^{n-1})\prod_{j=3}^4(1-a_1a_jq^n)}
  {a_1(1-b_4q^{2n-1})(1-b_4q^{2n})}
  +\frac{a_1a_3a_4q^{n-1}(1-q^n)\prod_{j=3}^4(1-a_2a_jq^{n-1})}
  {(1-b_4q^{2n-2})(1-b_4q^{2n-1})},\n
  &\quad\tilde{C}_n=\frac{-a_3a_4q^{n-1}(1-q^n)\prod_{j=1}^2\prod_{k=3}^4
  (1-a_ja_kq^{n-1})}
  {(1-b_4q^{2n-2})(1-b_4q^{2n-1})}.\nonumber
\end{align}

\section{Quantum Dilogarithm Function}
\label{app:qdilog}

For readers' convenience, we present some properties of the quantum dilogarithm
function (see \cite{qdilog,qdilog2,qdilog3} and references therein),
which are needed for calculations in the text.
There are several definitions for `quantum dilogarithm function' in the
literature. For example, the Faddeev's quantum dilogarithm given in
\cite{qdilog3} is defined by ($b\in\mathbb{C}$, $\text{Re}\,b\neq 0$),
\begin{equation}
  \Phi^{\text{F}}_b(z)=\exp\Bigl(-\int_{\mathbb{R}+i0}
  \frac{e^{-2izt}}{4\sinh bt\,\sinh b^{-1}t}\frac{dt}{t}\Bigr)\quad
  \bigl(|\text{Im}\,z|<\bigl|\tfrac{1}{2}\,\text{Re}\,(b+b^{-1})\bigr|\bigr).
  \label{Fqdilog}
\end{equation}
By analytic continuation, it is defined on the entire $z$-plane.
Our quantum dilogarithm $\Phi_{\gamma}(z)$ given below is
$\Phi_{\gamma}(z)=\Phi^{\text{F}}_{\!\!\!\sqrt{\frac{\pi}{\gamma}}}
\bigl(\frac{z}{2\sqrt{\pi\gamma}}\bigr)^{-1}$.
We assume $\gamma>0$ in this Appendix.
\\[2pt]\indent
Our definition is the following.
The quantum dilogarithm function $\Phi_{\gamma}(z)$ is a meromorphic function,
which is defined for $|\text{Im}\,z|<\gamma+\pi$ by the integral
representation \eqref{intrep} and analytically continued to the whole
complex plane by the functional equation \eqref{funcrel}.

\medskip
\noindent
\underline{definition}:
\begin{equation}
  \Phi_{\gamma}(z)=\exp\Bigl(\int_{\mathbb{R}+i0}
  \frac{e^{-izt}}{4\sinh\gamma t\,\sinh\pi t}\frac{dt}{t}\Bigr)
  \qquad(|\text{Im}\,z|<\gamma+\pi).
  \label{intrep}
\end{equation}
By taking the integration contour as $(-\infty,-\rho)+C_{\rho}+(\rho,\infty)$,
where $C_{\rho}$ is a semicircle with radius $\rho$
($0<\rho<\min(\frac{\pi}{\gamma},1)$) in the upper half plane,
the integral can be evaluated as
\begin{equation}
  \int_{\mathbb{R}+i0}\frac{e^{-izt}}{4\sinh\gamma t\,\sinh\pi t}\frac{dt}{t}
  =\int_{\rho}^{\infty}\frac{\sin zt}{2i\sinh\gamma t\,\sinh\pi t}\frac{dt}{t}
  +\int_0^{\pi}\frac{e^{-iz\rho e^{i\theta}}}
  {4i\sinh\gamma\rho e^{i\theta}\,\sinh\pi\rho e^{i\theta}}d\theta,
  \label{intrep2}
\end{equation}
which is independent of the choice of $\rho$.

\medskip
\noindent
\underline{functional relations}:
\begin{align}
  &\frac{\Phi_{\gamma}(z+i\gamma)}{\Phi_{\gamma}(z-i\gamma)}
  =\frac{1}{1+e^z},
  \label{funcrel}\\
  &\frac{\Phi_{\gamma}(z+i\pi)}{\Phi_{\gamma}(z-i\pi)}
  =\frac{1}{1+e^{\frac{\pi}{\gamma}z}},\\
  &\Phi_{\gamma}^*(z)=\frac{1}{\Phi_{\gamma}(z)},\\
  &\Phi_{\gamma}(z)\Phi_{\gamma}(-z)=
  \exp\Bigl(\frac{i}{4\gamma}\bigl(z^2+\frac{\gamma^2+\pi^2}{3}\bigl)\Bigr).
\end{align}
\underline{poles and zeros}:
\begin{align}
  \text{poles of $\Phi_{\gamma}(z)$} &:
  z=i\bigl((2n_1-1)\gamma+(2n_2-1)\pi\bigr)\quad
  (n_1,n_2\in\mathbb{Z}_{\geq 1}),
  \label{pole}\\
  \text{zeros of $\Phi_{\gamma}(z)$} &:
  z=-i\bigl((2n_1-1)\gamma+(2n_2-1)\pi\bigr)\quad
  (n_1,n_2\in\mathbb{Z}_{\geq 1}).
  \label{zero}
\end{align}
For $\frac{\gamma}{\pi}\not\in\mathbb{Q}$,
these poles \eqref{pole} and zeros \eqref{zero} are simple.\\
\underline{series expansions}: ($|\text{Im}\,z|<\gamma+\pi$)
\begin{align}
  \Phi_{\gamma}(z)&=
  \begin{cases}
  {\displaystyle
  \exp\biggl(
  \frac{i}{4\gamma}\Bigl(z^2+\frac{\gamma^2+\pi^2}{3}\Bigr)
  +i\sum_{n=1}^{\infty}\frac{(-1)^n}{2n}\Bigl(
  \frac{e^{-\frac{\pi z}{\gamma}n}}{\sin\frac{\pi^2n}{\gamma}}
  +\frac{e^{-zn}}{\sin\gamma n}\Bigr)\!\biggr)}
  &\!\!(\text{Re}\,z>0)\\
  {\displaystyle
  \exp\biggl(
  -i\sum_{n=1}^{\infty}\frac{(-1)^n}{2n}\Bigl(
  \frac{e^{\frac{\pi z}{\gamma}n}}{\sin\frac{\pi^2n}{\gamma}}
  +\frac{e^{zn}}{\sin\gamma n}\Bigr)\!\biggr)}
  &\!\!(\text{Re}\,z<0)\\
  {\displaystyle
  \lim_{\varepsilon\to+0}\Phi_{\gamma}(\pm\varepsilon+z)}
  &\!\!(\text{Re}\,z=0)
  \end{cases}\n
  &\quad\ \text{for $\tfrac{\gamma}{\pi}\not\in\mathbb{Q}$},
  \label{series}\\
  \Phi_{\gamma}(z)&=
  \begin{cases}
  {\displaystyle
  \exp\biggl(
  \frac{i}{4\gamma}\Bigl(z^2+\frac{\gamma^2+\pi^2}{3}\Bigr)
  +i\!\!\!\!\!
  \sum_{\genfrac{}{}{0pt}{}{n=1}{n\not\equiv0\,(\text{mod}\,M)}}^{\infty}
  \!\!\!\!\!\!\!\frac{(-1)^n}{2n}
  \frac{e^{-\frac{\pi z}{\gamma}n}}{\sin\frac{\pi^2n}{\gamma}}
  +i\!\!\!\!\!
  \sum_{\genfrac{}{}{0pt}{}{n=1}{n\not\equiv0\,(\text{mod}\,N)}}^{\infty}
  \!\!\!\!\!\!\!\frac{(-1)^n}{2n}
  \frac{e^{-zn}}{\sin\gamma n}\biggr)}
  &\!\!(\text{Re}\,z>0)\\
  {\displaystyle
  \exp\biggl(
  -i\!\!\!\!\!
  \sum_{\genfrac{}{}{0pt}{}{n=1}{n\not\equiv0\,(\text{mod}\,M)}}^{\infty}
  \!\!\!\!\!\!\!\frac{(-1)^n}{2n}
  \frac{e^{\frac{\pi z}{\gamma}n}}{\sin\frac{\pi^2n}{\gamma}}
  -i\!\!\!\!\!
  \sum_{\genfrac{}{}{0pt}{}{n=1}{n\not\equiv0\,(\text{mod}\,N)}}^{\infty}
  \!\!\!\!\!\!\!\frac{(-1)^n}{2n}
  \frac{e^{zn}}{\sin\gamma n}\biggr)}
  &\!\!(\text{Re}\,z<0)\\
  {\displaystyle
  \lim_{\varepsilon\to+0}\Phi_{\gamma}(\pm\varepsilon+z)}
  &\!\!(\text{Re}\,z=0)
  \end{cases}\n
  &\quad\ \text{for $\gamma=\tfrac{M}{N}\pi$ ($N$ and $M$ are positive
  integers and coprime)}.
  \label{seriesrat}
\end{align}
\underline{asymptotic forms}: ($|\text{Im}\,z|<\gamma+\pi$)
\begin{equation}
  \Phi_{\gamma}(z)\simeq
  \begin{cases}
  \exp\Bigl(\frac{i}{4\gamma}\bigl(z^2+\frac{\gamma^2+\pi^2}{3}\bigr)\Bigr)
  &\!\!(\text{Re}\,z\to\infty)\\[2pt]
  1&\!\!(\text{Re}\,z\to-\infty)
  \end{cases}.
  \label{asymp}
\end{equation}
\underline{$\gamma\to0$ limit}:
\begin{equation}
  \Phi_{\gamma}(z)=\exp\Bigl(\frac{1}{2i\gamma}\text{Li}_2(-e^z)
  +\text{O}(\gamma)\Bigr),
  \label{todilog}
\end{equation}
where the dilogarithm function $\text{Li}_2(z)$ is defined by
$\text{Li}_2(z)=\sum_{k=1}^{\infty}\frac{z^k}{k^2}$ ($|z|<1$) and
analytic continuation.

\bigskip

For $\Phi^{\text{F}}_b(z)$ in \eqref{Fqdilog}, we have\\
\underline{infinite product form of $\Phi^{\text{F}}_b(z)$}:
($\text{Im}\,b^2>0$)
\begin{equation}
  \Phi^{\text{F}}_b(z)
  =\frac{(-e^{2\pi b^{-1}z-\pi ib^{-2}};e^{-2\pi ib^{-2}})_{\infty}}
  {(-e^{2\pi bz+\pi ib^2};e^{2\pi ib^2})_{\infty}}.
  \label{Fqdilogprod}
\end{equation}

\goodbreak


\end{document}